\documentclass[fleqn,usenatbib]{mnras}

\usepackage[T1]{fontenc}

\DeclareRobustCommand{\VAN}[3]{#2}
\let\VANthebibliography\thebibliography
\def\thebibliography{\DeclareRobustCommand{\VAN}[3]{##3}\VANthebibliography}

\usepackage{graphicx}	% Including figure files
\usepackage{amsmath}	% Advanced maths commands
\usepackage{amssymb}	% Extra maths symbols

\usepackage{newtxtext,newtxmath}
\newcommand{\OIII}{[\ion{O}{III}]}
\newcommand{\Ha}{H$\alpha$}
\newcommand{\Hb}{H$\beta$}
\newcommand{\Hg}{H$\gamma$}
\newcommand{\Hd}{H$\delta$}
\newcommand{\He}{H$\epsilon$}
\newcommand{\NeIII}{[\ion{Ne}{III}]}
\newcommand{\OII}{[\ion{O}{II}]}

\title[When cosmic dawn breaks]{EPOCHS IX. When cosmic dawn breaks: Evidence for evolved stellar populations in $7 < z < 12$ galaxies from PEARLS GTO and public NIRCam imaging}

\author[J.\@A.\@A.\@ Trussler et al.]{James A.\@ A.\@ Trussler,$^{1}$\thanks{E-mail: james.trussler@manchester.ac.uk}
Christopher J. Conselice,$^{1}$
Nathan Adams,$^{1}$
Duncan Austin,$^{1}$
Leonardo Ferreira,$^{2}$
\newauthor
Tom Harvey,$^{1}$
Qiong Li,$^{1}$
Aswin P.\@ Vijayan,$^{3,4}$
Stephen M.\@ Wilkins,$^{5,6}$
Rogier A.\@ Windhorst,$^{7}$
\newauthor
Rachana Bhatawdekar,$^{8}$
Cheng Cheng,$^{9}$
Dan Coe,$^{10}$
Seth H.\@ Cohen,$^{7}$
Simon P.\@ Driver,$^{11}$
Brenda Frye,$^{12}$
\newauthor
Norman A.\@ Grogin,$^{10}$
Nimish Hathi,$^{10}$
Rolf A.\@ Jansen,$^{7}$
Anton Koekemoer,$^{10}$
Madeline A.\@ Marshall,$^{13,14}$
\newauthor
Mario Nonino,$^{15}$
Rafael Ortiz,$^{7}$
Nor Pirzkal,$^{10}$
Aaron Robotham,$^{11}$
Russell E.\@ Ryan, Jr.,\@$^{10}$
\newauthor
Jordan C.\@ J.\@ D'Silva,$^{11,14}$
Jake Summers,$^{7}$
Scott Tompkins,$^{7}$
Christopher N.\@ A.\@ Willmer$^{12}$
and Haojing Yan$^{16}$
\\
$^{1}$Jodrell Bank Centre for Astrophysics, University of Manchester, Oxford Road, Manchester M13 9PL, UK\\
$^{2}$Department of Physics \& Astronomy, University of Victoria, Finnerty Road, Victoria V8P 1A1, Canada\\
$^{3}$Cosmic Dawn Center (DAWN)\\ 
$^{4}$DTU-Space, Technical University of Denmark, Elektrovej 327, DK-2800 Kgs. Lyngby, Denmark\\
$^{5}$Astronomy Centre, University of Sussex, Falmer, Brighton BN1 9QH, UK\\
$^{6}$Institute of Space Sciences and Astronomy, University of Malta, Msida MSD 2080, Malta\\
$^{7}$School of Earth and Space Exploration, Arizona State University, Tempe, AZ 85287-1404, USA\\
$^{8}$European Space Agency (ESA), European Space Astronomy Centre (ESAC), Camino Bajo del Castillo s/n, 28692 Villanueva de la Cañada, Madrid, Spain\\
$^{9}$Chinese Academy of Sciences South America Center for Astronomy, National Astronomical Observatories, CAS, Beijing 100101, China\\
$^{10}$Space Telescope Science Institute, 3700 San Martin Drive, Baltimore, MD 21218, USA\\
$^{11}$International Centre for Radio Astronomy Research (ICRAR) and the
International Space Centre (ISC),\\The University of Western Australia, M468,
35 Stirling Highway, Crawley, WA 6009, Australia\\
$^{12}$Steward Observatory, University of Arizona, 933 N Cherry Ave, Tucson, AZ, 85721-0009, USA\\
$^{13}$National Research Council of Canada, Herzberg Astronomy \& Astrophysics Research Centre, 5071 West Saanich Road, Victoria, BC V9E 2E7, Canada\\
$^{14}$ARC Centre of Excellence for All Sky Astrophysics in 3 Dimensions (ASTRO 3D), Australia\\
$^{15}$INAF-Osservatorio Astronomico di Trieste, Via Bazzoni 2, 34124 Trieste, Italy\\
$^{16}$Department of Physics and Astronomy, University of Missouri, Columbia, MO 65211, USA
}

\date{Accepted XXX. Received YYY; in original form ZZZ}

\pubyear{2024}

\begin{document}
\label{firstpage}
\pagerange{\pageref{firstpage}--\pageref{lastpage}}
\maketitle

\begin{abstract}
The presence of evolved stars in high-redshift galaxies can place valuable indirect constraints on the onset of star formation in the Universe. Thus we use PEARLS GTO and public NIRCam photometric data to search for Balmer-break candidate galaxies at $7 < z < 12$. We find that our Balmer-break candidates at $z \sim 10.5$ tend to be older (115~Myr), have lower inferred \OIII\ + \Hb\ equivalent widths (120~\AA), have lower specific star formation rates (6~Gyr$^{-1}$) and redder UV slopes ($\beta = -1.8$) than our control sample of galaxies. However, these trends all become less strong at $z \sim 8$, where the F444W filter now probes the strong rest-frame optical emission lines, thus providing additional constraints on the current star formation activity of these galaxies. Indeed, the bursty nature of Epoch of Reionisation galaxies can lead to a disconnect between their current SED profiles and their more extended star formation histories. We discuss how strong emission lines, the cumulative effect of weak emission lines, dusty continua, and AGN can all contribute to the photometric excess seen in the rest-frame optical, thus mimicking the signature of a Balmer break. Additional medium-band imaging will thus be essential to more robustly identify Balmer-break galaxies. However, the Balmer break alone cannot serve as a definitive proxy for the stellar age of galaxies, being complexly dependent on the star-formation history. Ultimately, deep NIRSpec continuum spectroscopy and MIRI imaging will provide the strongest indirect constraints on the formation era of the first galaxies in the Universe, thereby revealing when cosmic dawn breaks.
\end{abstract}


%%%%%%%%%%%%%%%%%%%%%%%%%%%%%%%%%%%%%%%%%%%%%%%%%%

%%%%%%%%%%%%%%%%% BODY OF PAPER %%%%%%%%%%%%%%%%%%

\section{Introduction} \label{sec:intro}

The \emph{James Webb Space Telescope} \citep[\emph{JWST,}][]{Gardner2023} is forever changing our view of the cosmos. Among its chief scientific goals \citep{Gardner2006} is to witness the beginning, the emergence of the First Light, through observing the formation of the very first galaxies in the Universe, shortly after the Big Bang. Indeed that first emerging primordial starlight, the break of cosmic dawn, marks a key transition point: from a Universe shrouded in darkness, the `Cosmic Dark Ages', to a Universe bathed in light, the onset of the Epoch of Reionisation. 

Our search for the first stars and galaxies has a rich history \citep[see e.g.\@][]{Dunlop2013, Finkelstein2016}, being greatly accelerated in the \emph{Hubble} + \emph{Spitzer} era. With the incorporation of increasingly more sensitive imaging instrumentation, which image at increasingly longer wavelengths, the redshift frontier steadily increased, especially so with the introduction of \emph{Hubble's} Advanced Camera for Surveys (ACS) \citep[see e.g.\@][]{Bunker2003, Bouwens2004, Stanway2004, Yan2004} and Wide Field Camera 3 (WFC3) \citep[see e.g.][]{Bouwens2010, Bunker2010, Finkelstein2010, McLure2010, Oesch2010}. The sensitive, near-infrared (NIR) filters on these instruments enabled galaxies at increasingly higher redshifts, which begin to dropout \citep[due to Ly$\alpha$ attenuation by the intergalactic medium (IGM), see e.g.\@][]{Madau1995, Inoue2014} at increasingly longer wavelengths, to be identified, thus bringing us ever closer to the formation epoch of the first galaxies at cosmic dawn. However, this search for the highest redshift galaxies in the \emph{HST} + \emph{Spitzer} era hit a natural redshift frontier \cite[see e.g.\@][]{Bouwens2011, Bouwens2019, Coe2013, Ellis2013, Oesch2012, Oesch2018}. And this frontier was not because we had already reached the formation epoch of the very first galaxies in the Universe (which we hope to achieve in the \emph{JWST} era), but rather simply due to running out of sensitive NIR \emph{HST} filters with which to actually detect these high-redshift dropout galaxies (as well as the low resolution of \emph{Spitzer}). With WFC3/F160W, this placed the long-standing redshift frontier at $z{\sim}$10--11, with perhaps the best (i.e.\@ with a likely \emph{HST} grism detection of the Ly$\alpha$ break, see \citealt{Oesch2016}, and a possible detection of rest-frame UV emission lines, see \citealt{Jiang2021}) high-redshift candidate in the \emph{HST} + \emph{Spitzer} era being GN-z11 \citep{Oesch2016}, a bright $z\sim 11$ galaxy in the GOODS North field (though see e.g.\@ also \citealt{Zheng2012}, \citealt{Coe2013}, \citealt{Zitrin2014} for other $z{\sim}$10 candidates, with the \citealt{Ellis2013} $z=11.9$ candidate, identified from a single-band WFC3/F160W detection, now confirmed to be at $z=11.5$, see \citealt{Curtis-Lake2023}).  With \emph{JWST} spectroscopy, this galaxy has now been spectroscopically-confirmed to be at $z=10.6$ \citep{Bunker2023}, hosting a wealth of emission lines \citep{Cameron2023, Maiolino2023a} and a clearly detected Ly$\alpha$ break \citep{Tacchella2023b}.

Despite the $z \sim 10$ redshift frontier for identifying high-redshift candidates, valuable constraints on the formation of $z > 10$ galaxies were able to be placed in the \emph{HST} + \emph{Spitzer} era. Just as old stars in our Milky Way Galaxy enable us to wind back the cosmic clock beyond the epoch of observation, chronicling the formation history of the Galaxy billions of years into the cosmic past \citep[see e.g.\@][]{Tolstoy2009, Snaith2014, Frebel2015, Xiang2022}, so too do the presence of old stars in high-redshift galaxies, when the Universe was very young, enable us to place valuable indirect constraints on the onset of star formation in the Universe. This indirect search for the first starlight, too, has a rich history. With, for example, the identification of a $\sim$3.5~Gyr old galaxy at $z=1.55$ \citep[when the Universe was 4.15~Gyr old,][]{Dunlop1996, Spinrad1997}, galaxies $\sim$800~Myr old at $z\sim 3.75$ \citep[when the Universe was 1.65~Gyr old,][]{Nayyeri2014}, galaxies several hundred Myr old at $z\sim6$ \citep[when the Universe was 930~Myr old,][]{Mobasher2005, Yan2006} and most recently  with six galaxies inferred to be $> 250$~Myr old at $z\sim 9$ \citep[when the Universe was 540~Myr old,][]{Laporte2021}. 

Indeed, in the epoch of reionisation (EoR), the most readily identifiable and accessible indicator of evolved stellar populations in the \emph{HST} + \emph{Spitzer} (and now \emph{JWST}) bands is the Balmer break, a sudden jump in the continuum level at 3646~\AA\ rest-frame, which manifests itself as a photometric excess in the 3.6~\textmu m (at $z \sim 8$) and 4.5~\textmu m (at $z \sim 10$) \emph{Spitzer}/IRAC bands \citep[see e.g.\@][]{Roberts-Borsani2020, Laporte2021}. Serving as an indicator for intermediately aged stellar populations (several hundred Myr), the Balmer break is attributable to A-type stars, which are sufficiently hot such that the majority of the hydrogen gas in their atmospheres is in the $n=2$ state, but not too hot that the hydrogen gas is ionised (such as for the hotter, brighter, shorter-lived O- and B-type stars). These stars are therefore very effective at absorbing their own stellar radiation at wavelengths capable of shifting their electrons from the $n=2$ state to a higher energy level ($n \geq 3$), resulting in Balmer absorption lines \citep[for an early analysis of galaxy stellar ages, see e.g.\@][]{Couch1987}, as well as radiation that is capable of ionising ($n = \infty$) the $n=2$ electrons, resulting in a strong attenuation of light more energetic than the Balmer limit (i.e.\@ $\lambda < 3646$~\AA) and therefore an increased (decreased) continuum level redward (blueward) of the resulting Balmer break. We note that the 4000~\AA\ break also serves as a stellar age indicator, though this builds up over longer (compared to the Balmer break), $\sim$Gyr timescales \citep[see e.g.\@][]{Kauffmann2003, Gallazzi2005}, and thus is likely negligible for galaxies in the EoR.

However, galaxies in the epoch of reionisation have also been thought \citep[and now confirmed by \emph{JWST}, see e.g.\@][]{Curti2023, Fujimoto2023, Matthee2023, Sanders2023, Tacchella2023a, Trussler2023} to be highly star-forming, with the strong (i.e.\@ high equivalent width) emission from the \OIII\ + \Hb\ rest-frame optical lines possibly contributing substantially to the bandpass-averaged flux density measured in broadband photometry \citep[see e.g.\@][]{Smit2014, Smit2015, Roberts-Borsani2016, Endsley2021}, boosting the observed fluxes well above the continuum level and thus mimicking the photometric excess signature of a Balmer break \citep[][]{Roberts-Borsani2020, Laporte2021}. Thus, in order to distinguish between Balmer breaks and strong line emission from photometric data, \citet{Laporte2021} required their Balmer-break candidate galaxies to be at sufficiently high redshift ($z > 9$) such that the strong rest-frame optical emission lines are redshifted out of the filter exhibiting the photometric excess (IRAC 4.5~\textmu m). In principle, this redshift cut implies the observed IRAC excess cannot be attributable to strong line emission, but is instead (more likely to be) due to a Balmer break. In this way, \citet{Laporte2021} derive stellar ages and star formation histories for their Balmer-break candidates, with a significant fraction (70\%) of stellar mass forming before $z=10$.

\emph{JWST}, with its unique set of imaging and spectroscopic capabilities in the near- and mid-infrared, is already beginning to transform our understanding of the formation and evolution of the first galaxies in the Universe. Indeed, in terms of the direct search for high-redshift galaxies, the long-standing $z \sim 10$ frontier has already been broken, with the identification of e.g.\@ $z \sim 12$ candidate galaxies from the very first Near Infrared Camera (NIRCam) data \citep{Pontoppidan2022, Treu2022, Finkelstein2023}, as reported in analyses \citep{Castellano2022, Finkelstein2022, Naidu2022, Adams2023, Atek2023, Donnan2023} just weeks after the commencement of scientific operations with \emph{JWST}. Furthermore, as of the time of writing this article, the highest spectroscopically-confirmed galaxy is at $z=13.2$ \citep{Curtis-Lake2023}, i.e.\@ 320~Myr after the Big Bang, already pushing 100~Myr earlier in cosmic history than the limits of \emph{HST} and \emph{Spitzer} \citep[GN-z11,][]{Oesch2016}. 

The transformative capabilities of \emph{JWST} are also beneficial to indirect analyses on the onset of star formation in the Universe, such as the one conducted in this work. Of course, for a given inferred stellar age, a galaxy detected at higher redshift naturally places stronger constraints on the formation era of the first galaxies. Furthermore, the great sensitivity of \emph{JWST} enables the fainter and much more numerous galaxy population to be uncovered, which may display different star formation histories to their brighter counterparts. Indeed, sources that once took several hundred hours to detect with \emph{Spitzer}/IRAC \citep[see e.g.\@][]{Stefanon2021b, Stefanon2021} and are well-beyond the capabilities of existing K-band imaging from the ground, can now be detected with \emph{JWST} in a matter of minutes. Hence much larger samples, of high S/N Balmer-break candidates can now be assembled. Perhaps most importantly however, are the new photometric bands that \emph{JWST} brings, which will allow for a much clearer characterisation of the physical properties of these high-redshift galaxies. This is partially due to the new F277W filter, but especially through the extensive set of medium-band filters \citep[see e.g.\@][]{Roberts-Borsani2021}, which allow for the most reliable identification of a Balmer break, as the separate contributions from emission lines and the underlying continuum can be effectively disentangled \citep[see e.g.\@][]{Laporte2023}

In this work we therefore make use of these excellent new capabilities of \emph{JWST}, utilising a combination of NIRCam imaging from the Prime Extragalactic Areas for Reionisation and Lensing Science Guaranteed Time Observations (PEARLS GTO) programme \citep{Windhorst2023} and public surveys (CEERS, GLASS and NGDEEP, see \citealt{Bagley2023a}, \citealt{Treu2022} and \citealt{Bagley2023b}), to search for Balmer-break galaxy candidates at $7 < z < 12$. More specifically, we apply colour cuts, which probe the strength of the (supposed) Balmer break, to identify candidates at $z \sim 10.5$ with a F444W excess, as well as (thanks to the new F277W filter) candidates at $z \sim 8$ with a F356W excess. Using the full available wide-band NIRCam photometry (and the medium-band F410M when available), we aim to determine the stellar ages and star formation histories of these Balmer-break galaxies, to (in principle) place new indirect constraints on the onset of star formation in the Universe. However, as we will discuss in detail in this paper, additional medium-band imaging will be essential to reliably identify Balmer breaks and thus better constrain the star formation histories in these high-redshift galaxies. Indeed, both strong emission lines, weak emission lines, dusty continua and a (dusty) AGN can all contribute to the photometric excess seen in our wide-band NIRCam filters, thus mimicking the signature of a Balmer break. With these degeneracies in mind, we also comment on the (extremely) large stellar masses ($\log\, (M_*/\mathrm{M}_\odot) > 10$) inferred for some red rest-frame optical Balmer-break candidate galaxies at $z \sim 8$ \citep{Labbe2023}. Finally, we discuss the limitations of the Balmer break strength as a proxy for the stellar age of a galaxy, advocating for deep follow-up NIRSpec continuum spectroscopy and MIRI imaging to derive the best possible star-formation history constraints for these galaxies. 

This paper is structured as follows. In Section~\ref{sec:data}, we discuss the NIRCam data which underpins our analysis, the methodology for constructing the EPOCHS parent sample of high-redshift galaxies (see \citealt{Adams2023} and Conselice et al.\@, in prep.), as well as our use of the Bagpipes \citep{Carnall2018} SED-fitting code to place constraints on the star formation properties of these galaxies. In Section~\ref{sec:selection}, we discuss the redshift and colour cuts we apply to identify Balmer-break candidates. In Section~\ref{sec:results}, we discuss the stellar ages, star formation histories and other physical properties inferred for these high-redshift galaxies. We also comment on the magnitudes (i.e.\@ brightnesses) and stellar masses of the Balmer-break candidates to establish what is likely driving the (extremely) large stellar masses inferred in the literature for some of these systems. In Section~\ref{sec:discussion} we discuss caveats regarding the identification of Balmer breaks from photometric data, as well as the constraints these Balmer breaks can place on the presence of evolved stars in high-redshift galaxies. Finally, in Section~\ref{sec:conclusions} we conclude. We use the AB magnitude system throughout and assume a \citet{Planck2020} cosmology.

\section{Data} \label{sec:data}

In this section we describe the NIRCam data underpinning our analysis, the methodology for constructing the EPOCHS high-redshift galaxy parent sample (see \citet{Adams2023} and Conselice et al.\@, in prep.), as well as the procedure for using the SED-fitting code Bagpipes \citep{Carnall2018} to derive constraints on the stellar ages, star formation histories and star formation properties of these high-redshift galaxies.

\subsection{NIRCam data}

We use a combination of PEARLS GTO \citep{Windhorst2023} and public NIRCam \citep{Rieke2005} imaging data \citep{Treu2022, Bagley2023a, Bagley2023b}. For full details of these datasets and the reduction process, we refer the reader to \citet{Adams2023} and Conselice et al.\@ (in prep.). Briefly, we use a modified version of the \emph{JWST} official pipeline, utilising pipeline version 1.8.2 and Calibration Reference Data System (CRDS) context jwst\_0995.pmap, which includes wisp subtraction, 1/f noise correction, background subtraction and alignment onto GAIA WCS. We briefly outline the key aspects of the NIRCam data we use in our analysis below.

\subsubsection{PEARLS imaging}

The PEARLS \citep[Prime Extragalactic Areas for Reionization and Lensing Science,][]{Windhorst2023} GTO is a 110~hour Cycle 1 programme, which is primarily obtaining medium-deep ($\sim$28--29~AB~mag depth) NIRCam imaging across a variety of separate extragalactic fields. These fields can be divided into two distinct classes: blank fields and cluster fields. 

In terms of the blank fields, we make use of the 4 pointings (spokes) obtained in the North Ecliptic Pole Time Domain Field \citep[NEP-TDF,][]{Jansen2018}, an extragalactic field that is always accessible year-round by \emph{JWST}. 

In terms of the cluster fields, we make use of the single pointing on El Gordo \citep[see e.g.\@][]{Diego2023, Frye2023}, together with the three epoch (pointing) data on the MACS 0416 cluster. In this work, in order to avoid having to apply magnification corrections/perform an intracluster light background subtraction, we do not make use of the data from the NIRCam module centred on the galaxy clusters themselves. Instead, we solely use the data from the other available NIRCam module, which is essentially imaging the blank field parallel (3~arcmin away) to the cluster itself.

8-band NIRCam imaging data is available for the three aforementioned PEARLS fields: F090W, F115W, F150W, F200W, F277W, F356W, F410M and F444W. 

\subsubsection{Public imaging}

We also make use of NIRCam imaging from a number of public surveys. Namely, CEERS, GLASS and NGDEEP. CEERS \citep{Bagley2023a} has 7 imaging bands (lacking F090W, compared to PEARLS), GLASS \citep{Treu2022} has 7 imaging bands (lacking F410M) and NGDEEP \citep{Bagley2023b} has 6 imaging bands (lacking F090W and F410M). We do not use the SMACS 0723 ERO data \citep{Pontoppidan2022} in this work, as the lack of F115W imaging can make it challenging  to properly distinguish between $z \sim 10$ Balmer-break galaxies and $z \sim 8$ strong \OIII\ + \Hb\ emitters \citep[see e.g.\@][]{Adams2023, Trussler2023}, thus being particularly detrimental for our analysis.

\subsection{EPOCHS high-redshift parent sample}

Here we briefly describe the methodology for constructing the EPOCHS parent sample of high-redshift galaxies. For full details we refer the reader to \citet{Adams2023} and Conselice et al.\@ (in prep.).

We use SExtractor \citep{Bertin1996} to identify sources in the F444W band. We then carry out forced aperture photometry across the various NIRCam bands, adopting a 0.32~arcsec diameter circular extraction aperture. The resulting flux densities are then aperture-corrected using corrections based off of the simulated WebbPSF \citep{Perrin2012, Perrin2014} point-spread functions for each band.

We fit the aperture-corrected 0.32~arcsec diameter aperture photometry using both the Le Phare \citep{Arnouts1999, Ilbert2006} and EAZY \citep{Brammer2008} SED-fitting codes. We note that for EAZY we use a combination of the default plus \citet{Larson2022} spectral templates. We impose a maximum SNR of 20 per NIRCam filter when SED-fitting the photometry, to account for remaining uncertainties in the photometric flux calibrations, as well as template mismatch between the observed data and the models. In order to increase the likelihood of selecting true high-redshift candidates, we apply the following quality cuts.

Firstly. we require the resulting redshift probability distribution functions to have an integrated probability $P > 0.8$ of the galaxy being at $z > 5$, both for Le Phare and for EAZY. Furthermore, the integrated probability within 10\% of the peak photometric redshift should be greater than 0.6. Additionally, we require the peak probability amplitude in this redshift range to be at least twice as large as the probability amplitude of the secondary, low-redshift solution (if present). Secondly, we require the galaxy to (begin to) drop out in one or more of the NIRCam bands. For fields with F090W imaging, this criterion translates into a rough lower redshift limit of $z\sim6.5$. For fields without F090W imaging, but that include F115W (i.e.\@ CEERS and NGDEEP), this criterion translates into a rough lower redshift limit of $z \sim 8.5$. Thirdly, given the assigned median photometric redshift, we require the candidate to be $> 5\sigma$ detected in the first two bands redward of the (supposed) Lyman break. We further require the candidate to be $< 3\sigma $ non-detected in the bands blueward of the Lyman break. This cut increases the confidence in the location of the (supposed) Lyman break, and reduces the chance of contamination by low-redshift interlopers (which typically have weaker Balmer/4000~\AA\ breaks). Fourth, we require a good quality of fit, i.e.\@ with both the Le Phare and EAZY reduced $\chi^2 < 6$. 

Following this procedure, we obtain the EPOCHS parent sample, which contains 204 $z > 6.5$ galaxy candidates.

\subsection{SED-fitting with Bagpipes}

We fit the photometric data with the SED-fitting code Bagpipes \citep{Carnall2018} to derive constraints on the star formation histories of these galaxies. As with Le Phare and EAZY, we set a minimum 5\% error (i.e. a maximum SNR of 20) per band. We use the default Bagpipes stellar and nebular templates, fitting the data assuming an exponential star-formation history.

We adopt uniform priors on the age (i.e.\@ time elapsed since onset of star formation) $0.001 < a\ (\mathrm{Gyr})$ < $\mathrm{Age_{Univ}}(z)$, $e$-folding time $0.1 < \tau (\mathrm{Gyr}) < 10$, stellar mass formed $1 < \log (M/\mathrm{M}_\odot) < 15$, metallicity $0 < Z/\mathrm{Z}_\odot < 2.5$, ionisation parameter $-4 < \log U < -2$ and V-band dust attenuation $0 < A_\mathrm{V} < 2$, assuming a \citet{Calzetti2000} dust attenuation law. Finally, following the methodology in our Austin et al.\@. (in prep.) companion paper, we adopt a Gaussian prior for the redshift, set by the mean and standard deviation of the EAZY redshift probability distribution function. The redshift is instead fixed at the known spectroscopic redshift when this is available. 

By default, Bagpipes outputs constraints on e.g.\@ the stellar mass, mass-weighted stellar age, star-formation history and specific star formation rate of the galaxy being fit. We make use of these quantities in our analysis.

Furthermore, we use the Bagpipes median-fit to derive the Balmer break strength (see definition in Section~\ref{sec:results}), \OIII\ doublet + \Hb\ equivalent width, absolute UV magnitude $M_{1500}$ and the UV slope $\beta$. We compute $M_{1500}$ from the median flux density over $1400 < \lambda_\mathrm{rest} $ (\AA) $< 1600$, while $\beta$ is computed by fitting a power-law over $1250 < \lambda_\mathrm{rest} $ (\AA) $ < 2600$ \citep[i.e.\@ the same wavelength range as in][]{Calzetti1994}. The errors on these quantities, for each galaxy, are determined by repeating the computation procedure for a set of 100 randomly selected posterior fits generated by Bagpipes, and determining the (16, 84) percentiles of the resulting quantity distributions. 

\section{Balmer break selection} \label{sec:selection}

\begin{figure*}
\centering
\includegraphics[width=\linewidth]{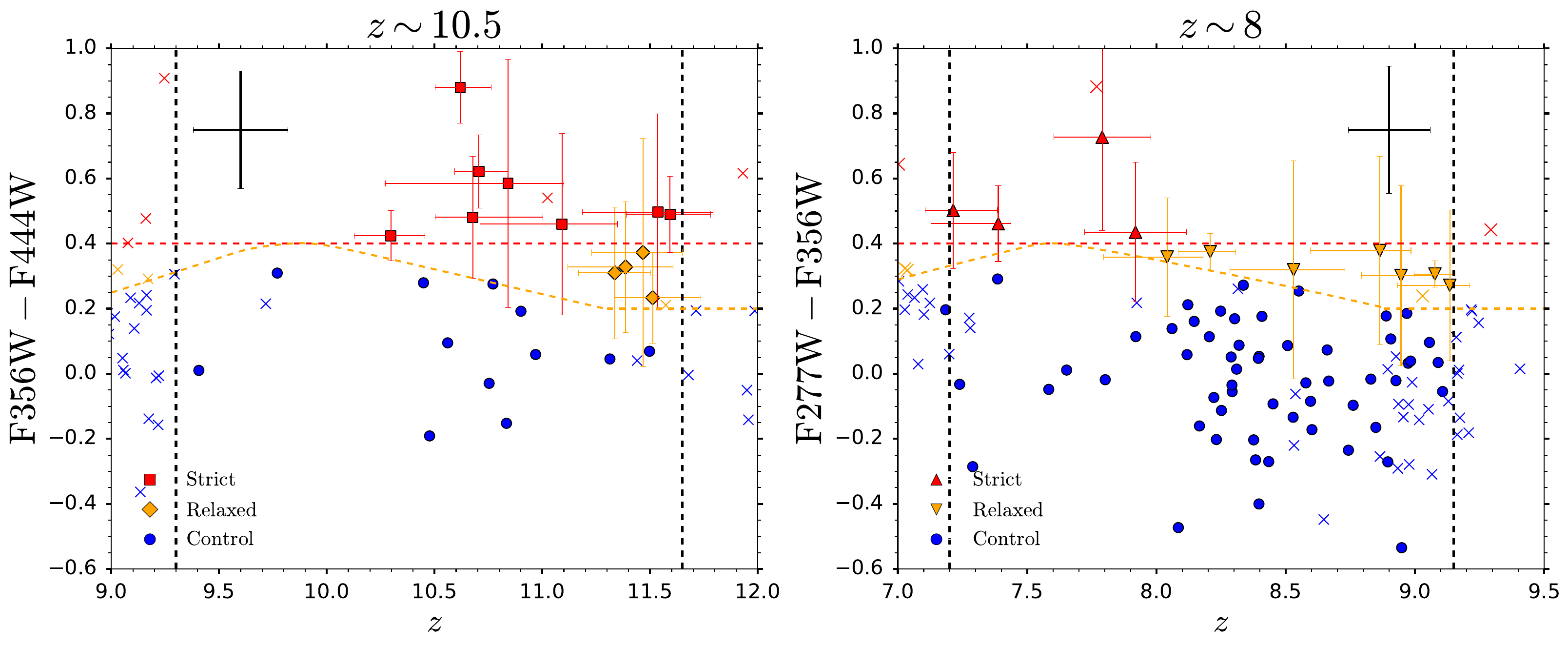}
\caption{Our colour--redshift selection for identifying Balmer-break candidates. In order to distinguish between line emission and Balmer breaks, we require the galaxies to be at sufficiently high redshift such that the \OIII\ and \Hb\ lines are redshifted out of the red filter in the filter pair defining the colour selection (left vertical dashed lines). We also require the galaxies to be at sufficiently low redshift such that the Balmer break is at most two-thirds of the way through the red filter (right vertical dashed lines). Our flat colour selection (red horizontal dashed lines) and redshift-dependent colour selection (orange) which accounts for the fact that the perceived strength of the Balmer break depends on where the break resides within the filter pair. Our strict (red squares/upward triangles), relaxed (orange diamonds/downward triangles) and control (blue circles) sample of galaxies. Crosses denote galaxies which do not satisfy our $>$ 68\% integrated redshift probability criterion. Individual error bars are shown for our strict and relaxed Balmer-break candidates. The median error bars for the full sample of galaxies are shown in the corner of each panel in black. Left panel: Our Balmer break selection at $z \sim 10.5$, for galaxies which exhibit a flux density excess in the F444W filter. Right panel: Our Balmer break selection at $z \sim 8$, for galaxies which exhibit a flux density excess in the F356W filter.}
\label{fig:balmer_break_selection}
\end{figure*}

In this section we describe the principles behind our selection of Balmer-break candidates, outlining the redshift and colour cuts we apply to generate our sample. We also show example SEDs and cutouts for both Balmer break and non-Balmer-break candidate galaxies. 

\subsection{Redshift cuts}

As described earlier, both strong line emission and Balmer breaks can give rise to elevated flux densities (i.e.\@ `excesses') in filters covering the rest-frame optical, potentially making it difficult to distinguish between these two effects \citep[see also][]{Roberts-Borsani2020, Laporte2021, Laporte2023}. Thus, in order to disentangle line emission from Balmer breaks, we follow the methodology first adopted by \citet{Laporte2021}, by requiring the galaxies to be at sufficiently high redshift such that the (potentially) strong line emission is redshifted out of the filter of interest (i.e.\@ the filter in which we wish to measure the photometric excess). Provided that the photometric redshift is correct, this then implies that the photometric excess observed is (mostly) attributable to a Balmer break, possibly providing evidence for evolved stellar populations in these high-redshift galaxies (though we will discuss various caveats regarding this procedure in Section~\ref{sec:discussion}).

More specifically, we require both the \OIII\ $\lambda$5007 and \Hb\ rest-frame optical lines to be redshifted out of the filter of interest (given the galaxy's photometric/spectroscopic redshift), as both of these lines can have very large equivalent widths \citep[up to $\sim$4000~\AA\ rest-frame for \OIII, see e.g.\@][]{Matthee2023} and thus can contribute substantially to the photometric excess seen in broadband photometry. Indeed, in contrast to \citet{Laporte2021}, we also require \Hb\ to be redshifted out as we have found \citep[see e.g.\@][]{Trussler2022} that the \Hb\ rest-frame equivalent width can in principle also be very high ($\sim$500~\AA) immediately ($<$ 3~Myr) after a starburst, producing a photometric excess $\Delta m$ ($=2.5\log_{10}[1 + \mathrm{EW_{rest}}(1+z)/\Delta \lambda ]$, where $\Delta \lambda$ is the filter width) of up to 0.4~mag. Even if such extremely strong \Hb\ line emission is not present, at the very least the \Hb\ emission can contribute a non-negligible amount of flux to the filter of interest, thereby reducing the accuracy with which the strength of the Balmer break can be determined. In our analysis we will focus on galaxies at $z\sim 10.5$ that exhibit a photometric excess in the F444W filter, as well as galaxies at $z\sim 8$ that first exhibit a photometric excess in the F356W filter. We note that the \OIII\ $\lambda$5007 and \Hb\ lines start to get redshifted out (i.e.\@ where the filter transmission drops to 50\% of its maximum) of the F444W filter at $z = 9.0, 9.3$, and out of the F356W filter at $z = 7.0, 7.2$, respectively. Thus the lower redshift limits for our F444W-excess and F356W-excess samples are $z=9.3$ and $z=7.2$, respectively.

In addition to a lower redshift limit, we also impose an upper redshift limit for our F444W- and F356W-excess samples, to account for the Balmer break itself redshifting out of the filter of interest. Indeed, just as the perceived photometric strength of the Lyman break depends on where the break resides within the wide-band filter of interest (being a maximum when the light blueward of the Lyman break fully spans a given filter and only $\sim$0.75~mag if it falls halfway through a filter), so too does the perceived strength of the Balmer break also depend on where the break resides within the filter. The perceived Balmer break strength (i.e.\@ the magnitude difference between the two neighbouring filters spanning the Balmer break) is at a maximum when the Balmer break is exactly between the two neighbouring filters, with the lower-lying continuum level blueward of the break fully in the bluer filter and the higher-lying continuum level redward of the break in the redder filter. The perceived Balmer break strength gradually weakens with increasing redshift as the lower-lying continuum blueward of the break begins to occupy an increasingly larger fraction of the redder filter in the filter pair, causing the bandpass-averaged flux density in this filter to decrease toward the lower continuum level. Hence the sensitivity to the Balmer break drops. To account for this, we therefore set an upper-redshift limit, which we deem to be when the Balmer break is two-thirds of the way through the filter of interest. This corresponds to $z=11.65$ and $z=9.15$ for the F444W and F356W filters, respectively. 

Hence when investigating galaxies with a potential Balmer break in the F444W filter, we restrict the redshift range to $9.30 < z < 11.65$ (i.e.\@ our $z\sim 10.5$ sample). For galaxies with a potential F356W excess, we instead restrict the redshift range to $7.20 < z < 9.15$ (i.e.\@ our $z\sim 8$ sample).

In order to increase the degree of confidence in the redshifts for our selected samples (and thus minimise contamination from e.g.\@ strong line emission), we require both the Le Phare and EAZY photometric redshift probability distributions to have an integrated probability greater than 68\% over the aforementioned redshift ranges.

In addition to the potential difficulty in distinguishing between Balmer-break galaxies and strong line emitters from photometry, there can also be general challenges in telling apart high-$z$ galaxies from low-$z$ interlopers, due to confusion between the Lyman and Balmer breaks. In principle the SEDs for these systems should be distinct, with the Lyman break strength essentially being infinite, while the Balmer break is finite. In practice however, due to the finite depth of observations, the SEDs of these systems can be comparable, especially for fainter systems close to the detection limit (e.g.\@ $5\sigma$ detected). Our high-redshift Balmer-break candidates should not suffer from this potential high--low redshift ambiguity, as by the nature of their selection, they all exhibit two prominent breaks in their photometry: the Lyman break and (supposedly) the Balmer break.

It is difficult to reproduce this distinct double break SED profile through a low-$z$ solution. Firstly, because this low-$z$ solution would have to have both a strong Balmer break (mimicking the Lyman break) and a second break (mimicking the high-$z$ Balmer break) seemingly powered by very strong line emission, which seems incompatible. Secondly, because the two breaks in the SED have to be at very specific wavelengths. The wavelength ratio between the Lyman-$\alpha$ break and the Balmer break is 3, while the ratio between the Lyman-$\alpha$ break and the \OIII+\Hb\ line complex is approximately 4. If the dual break SED is attributable to a low-$z$ Balmer break interloper, then its second break must be attributable to a prominent feature either at 1.09~\textmu m or 1.46~\textmu m rest-frame, of which there are none. Hence we can be confident in the high-redshift nature of our Balmer-break candidates.

\subsection{Colour cuts}

In addition to a redshift cut, we also apply a colour cut. Since we are interested in identifying potential Balmer break systems, we aim to find galaxies which have an excess in F444W (for $z \sim 10.5$ galaxies) or an excess in F356W (for $z \sim 8$ galaxies). This excess requirement translates into a sufficiently red F356W$-$F444W ($=C_{34}$) or F277W$-$F356W ($=C_{23}$) colour $C$. We consider two different colours cuts in our analysis. 

Firstly, we adopt a flat colour cut, requiring the colour $C > 0.4$~mag (i.e.\@ $C_{34} =$ F356W$-$F444W $>$ 0.4~mag at $z\sim 10.5$, or $C_{23} =$ F277W$-$F356W $>$ 0.4~mag at $z\sim 8$). The galaxies that satisfy this colour cut will be referred to as our strict sample. The motivation behind adopting the 0.4~mag threshold is firstly to reduce the chance of the colour selection being driven purely by noise, as if e.g.\@ both the F356W and F444W fluxes are measured at 5$\sigma$, the magnitude uncertainty in each filter is $\approx 0.2$~mag, so the F356W flux would have to be scattered down by $1\sigma$ and the F444W flux would have to be scattered up by $1\sigma$ to spuriously satisfy our colour requirement. Secondly, this threshold also reduces the chances of the red colour being driven by something other than a Balmer break (e.g.\@ dust or cumulative weak line emission, see Section~\ref{sec:discussion}). However, we acknowledge that the colour threshold is somewhat arbitrary. As we will discuss in more detail in Section~\ref{sec:discussion}, the greater the colour threshold that is adopted, typically the greater the age of the galaxies that are selected. 

Secondly, we adopt a redshift-dependent colour cut, that accounts for the fact that the perceived strength of the Balmer break depends on where the break resides within the filter pair comprising the colour selection. The galaxies that satisfy this colour cut (assuming the EAZY redshift), but not the flat colour cut, will be referred to as our relaxed sample.

The redshift-dependent colour threshold was determined by assuming that the Balmer break in the galaxy SED is given by a step function, which is comprised of a flat (in terms of $f_\nu$) continuum blueward of the break, and another flat continuum that is 0.4~mags brighter redward of the break. In other words, we assume the Balmer break has an intrinsic (i.e.\@ rest-frame) Balmer break strength of 0.4~mag. We then determine, as a function of redshift, the perceived strength (i.e.\@ observed-frame) of this Balmer break, namely the colour $C$ in F356W$-$F444W (for $z \sim 10.5$) and F277W$-$F356W (for $z \sim 8$). As shown in Fig.~\ref{fig:balmer_break_selection}, this redshift-dependent colour selection (orange) thus matches our flat colour selection (red) when the Balmer break is exactly in between the F356W and F444W filters (i.e.\@ $z=9.8$) or between the F277W and F356W filters (i.e.\@ $z=7.6$). However, the colour requirement becomes smaller as the redshift deviates from these values. Indeed, the main drawback of our flat colour selection is that a given observed colour $C$ demands an increasingly large intrinsic Balmer break strength as one deviates from the aforementioned redshifts. For reference, the $C = 0.4$~mag criterion requires an intrinsic Balmer break strength of 0.7~mags when the break is halfway through either the blue or red filter. The drawback of our redshift-dependent colour cut is that the smaller colour requirement means that we are more likely to select spurious objects. It is for this reason that we set a minimum colour threshold $C$ of 0.2~mags.

Finally, galaxies that satisfy neither our flat colour selection, nor our redshift-dependent colour selection, but are in the appropriate redshift range (i.e.\@ $9.30 < z < 11.65$ or $7.20 < z < 9.15$ for the $z \sim 10.5$ and $z \sim 8$ samples respectively), will be referred to as the control sample. In principle these should be galaxies with weak Balmer breaks and thus, on average, younger stellar populations. 

\subsection{Samples}

\begin{figure*}
\centering
\includegraphics[width=.475\linewidth] {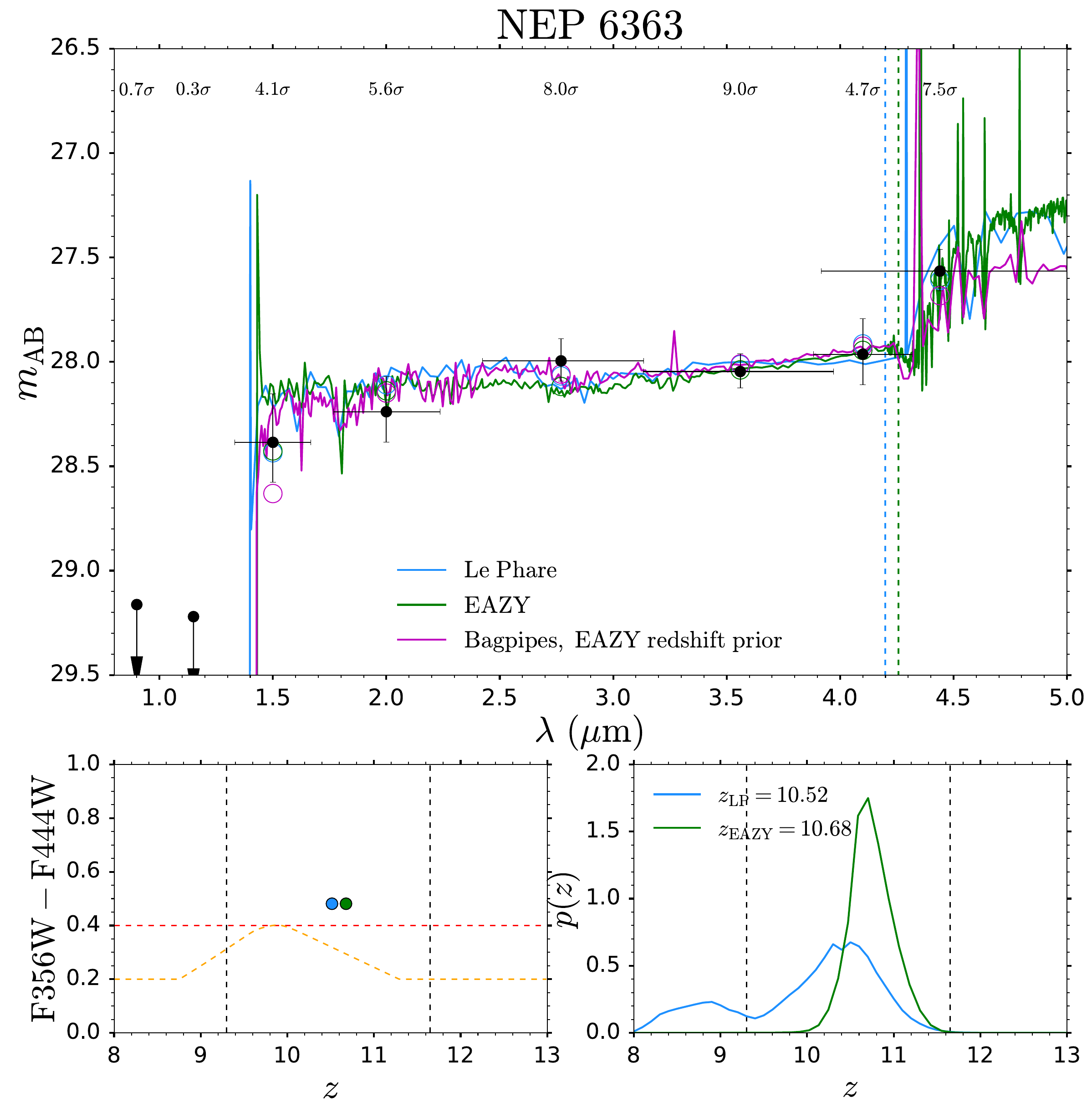} \hfill
\includegraphics[width=.475\linewidth]{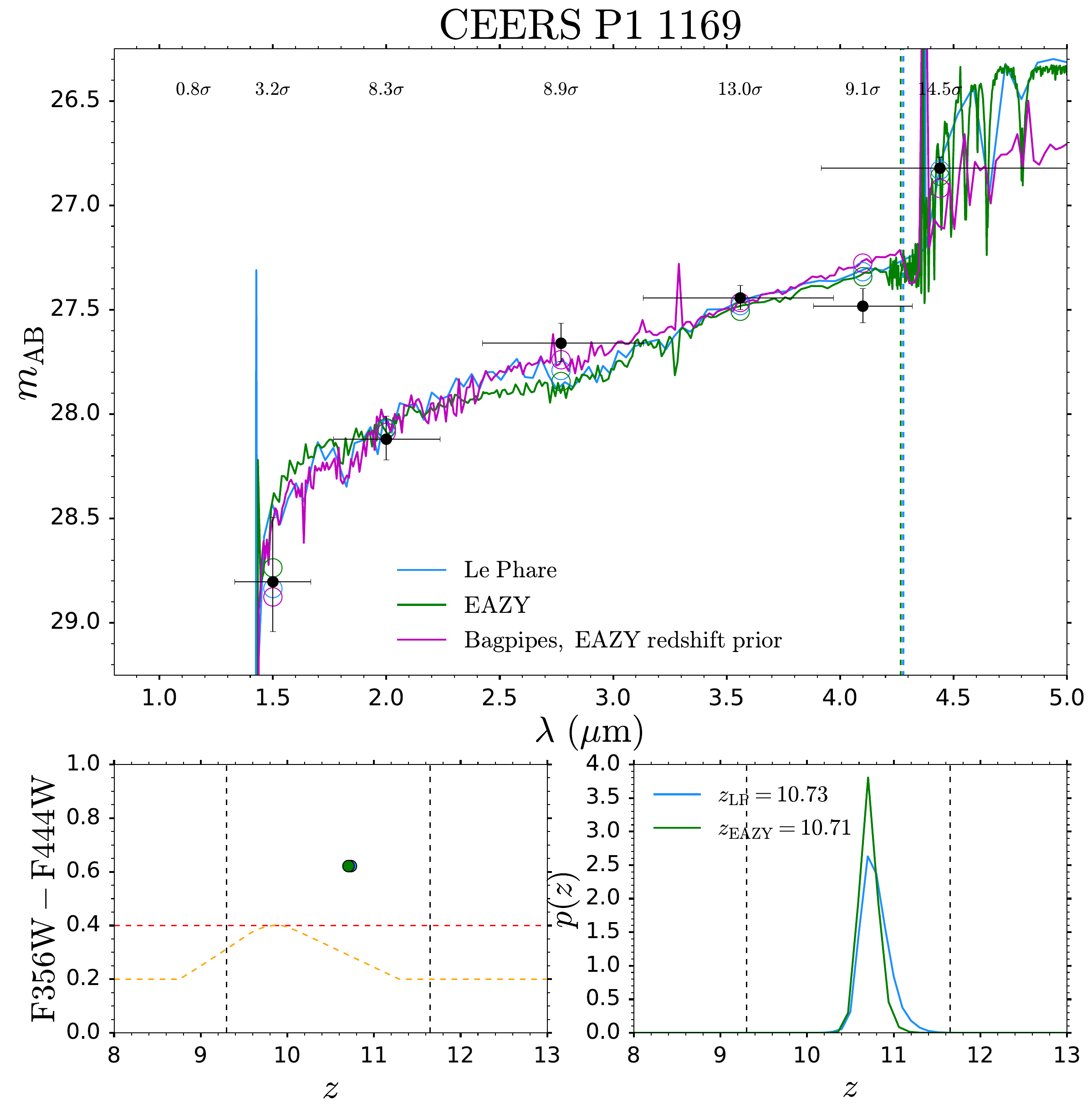} \\[4.5ex]
\includegraphics[width=.475\linewidth] {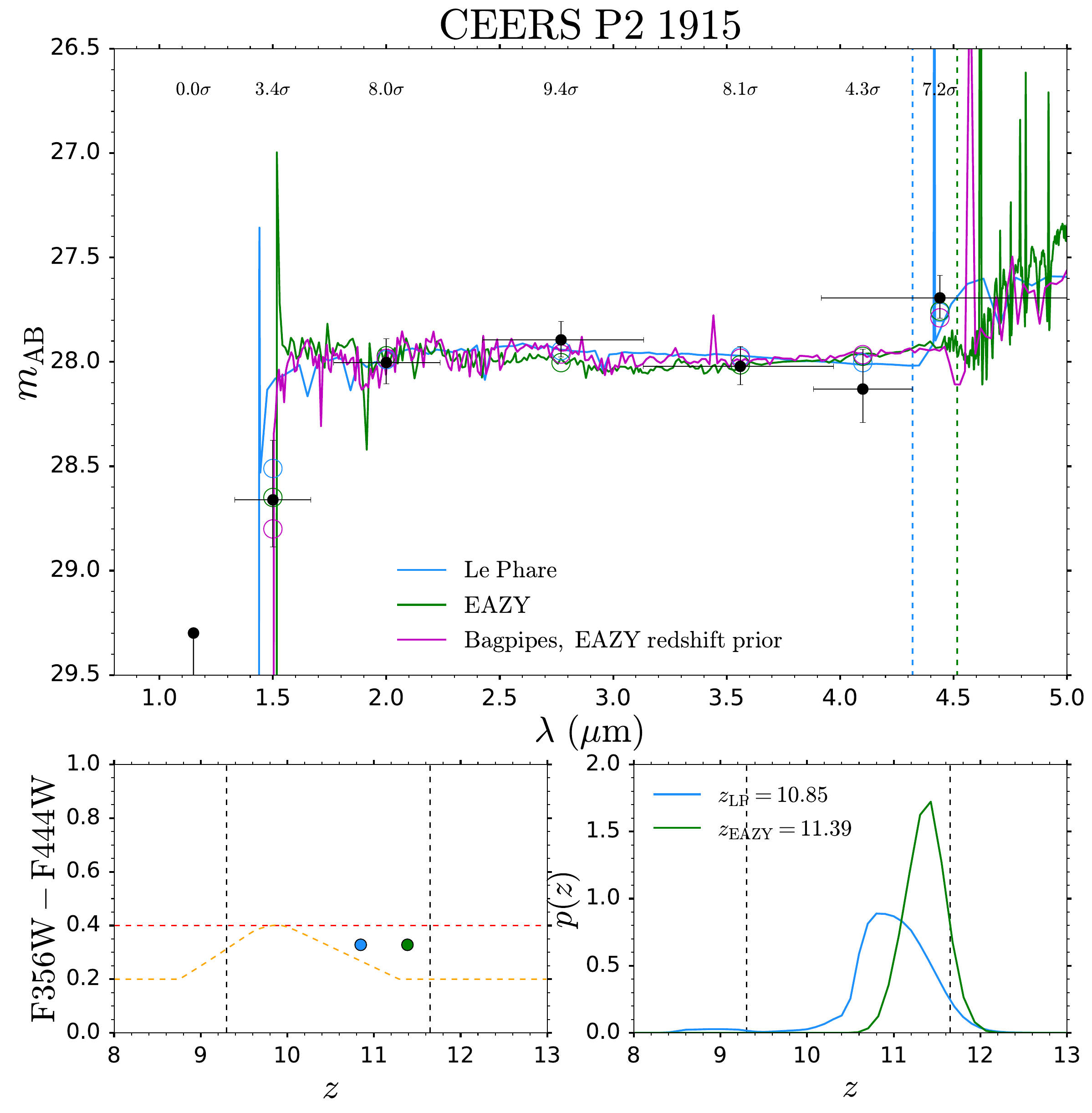} \hfill
\includegraphics[width=.475\linewidth]{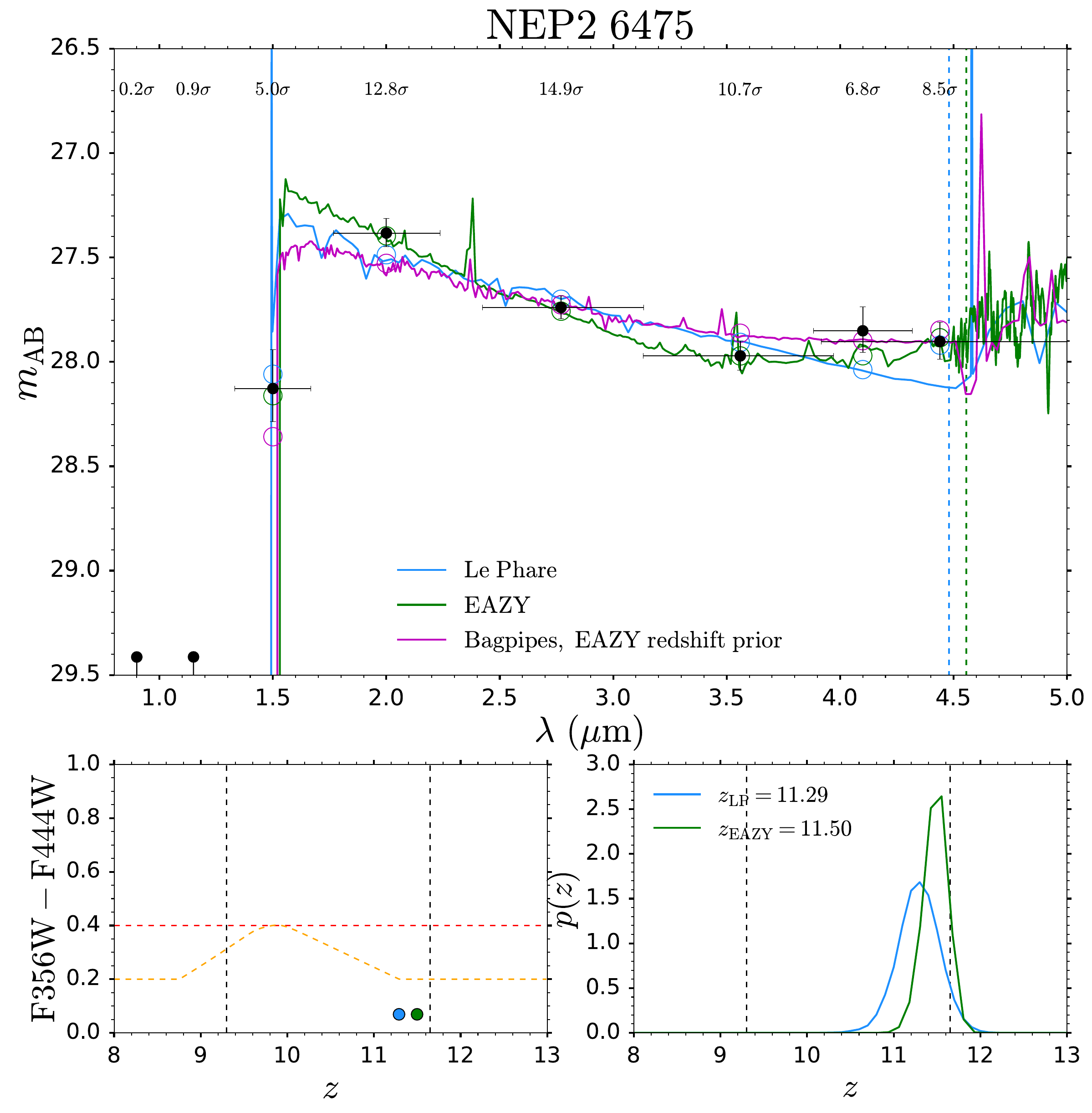} \\
\caption{Examples of galaxies that satisfy our strict Balmer break colour criterion and redshift requirement (top-left and top-right panels), a galaxy that satisfies our relaxed colour criterion (bottom-left) and a galaxy that does not satisfy either colour criterion (i.e.\@ the control sample, bottom-right), for our $z \sim 10.5$ sample. Top subpanels: NIRCam photometry (black) and SNR for each band (non-detections are displayed with downward arrows at the 2$\sigma$ depth), together with the best-fit Le Phare (light blue) and EAZY (green) SEDs, as well as the median-fit Bagpipes SED (magenta, which has been fit assuming either the EAZY redshift PDF as a prior, or fixed at the known spectroscopic redshift). The locations of the Balmer breaks (3646~\AA\ rest-frame) for the Le Phare and EAZY fits are denoted by the colour-coded dashed vertical lines. Bottom subpanels: our colour--redshift selection, as well as the Le Phare and EAZY photometric redshift probability distribution functions, with median redshifts displayed.} 
\label{fig:f444w_candidates}
\end{figure*}

We show our redshift- and colour-selections applied to our set of $z \sim 10.5$ and $z \sim 8$ galaxies in the left and right panels of Fig.~\ref{fig:balmer_break_selection}, respectively. The vertical dashed lines represent the lower and upper limits of our redshift selection. The flat red horizontal dashed line represents our fixed colour cut, while the orange curve represents our redshift-dependent colour criterion. The red, orange and blue data points corresponds to our strict, relaxed and control samples respectively. The F444W-excess (F356W-excess) Balmer-break candidates are shown in squares and diamonds (upward and downward triangles), control galaxies are shown as circles. The points denoted by crosses fail to satisfy our integrated redshift probability $>$ 68\% criterion. 

\begin{figure*}
\centering
\includegraphics[width=.475\linewidth] {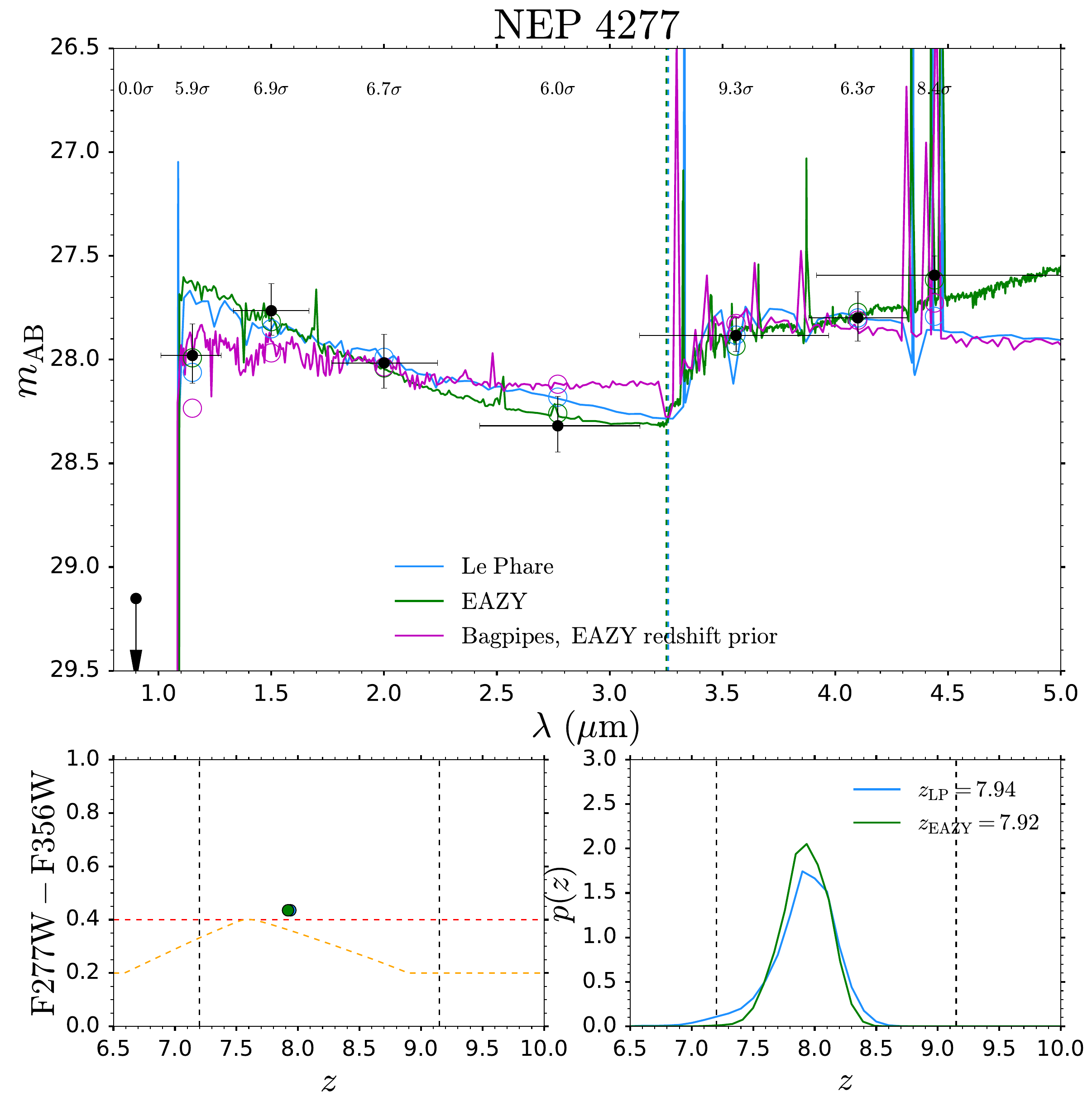} \hfill
\includegraphics[width=.475\linewidth]{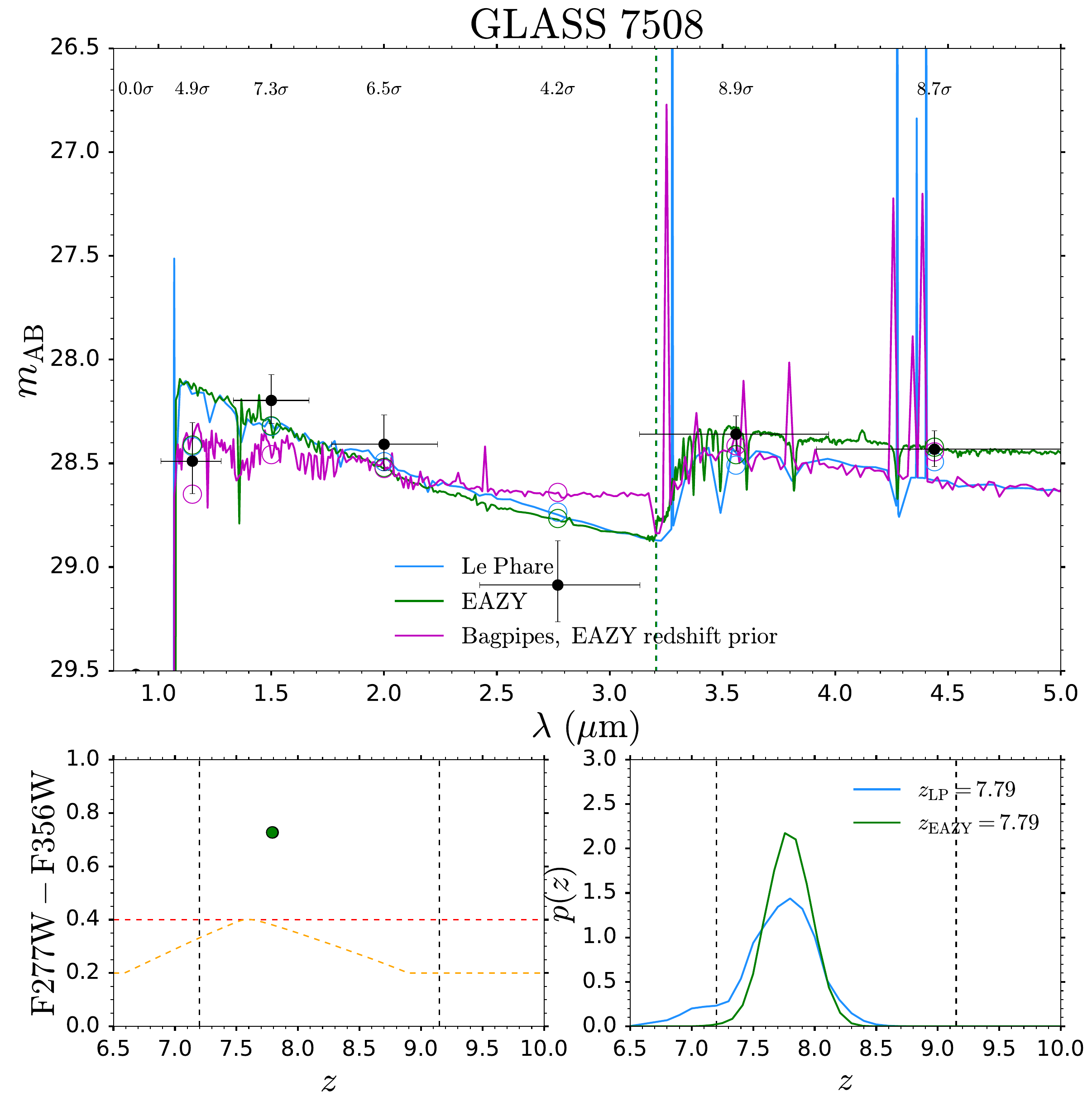} \\[4.5ex]
\includegraphics[width=.475\linewidth] {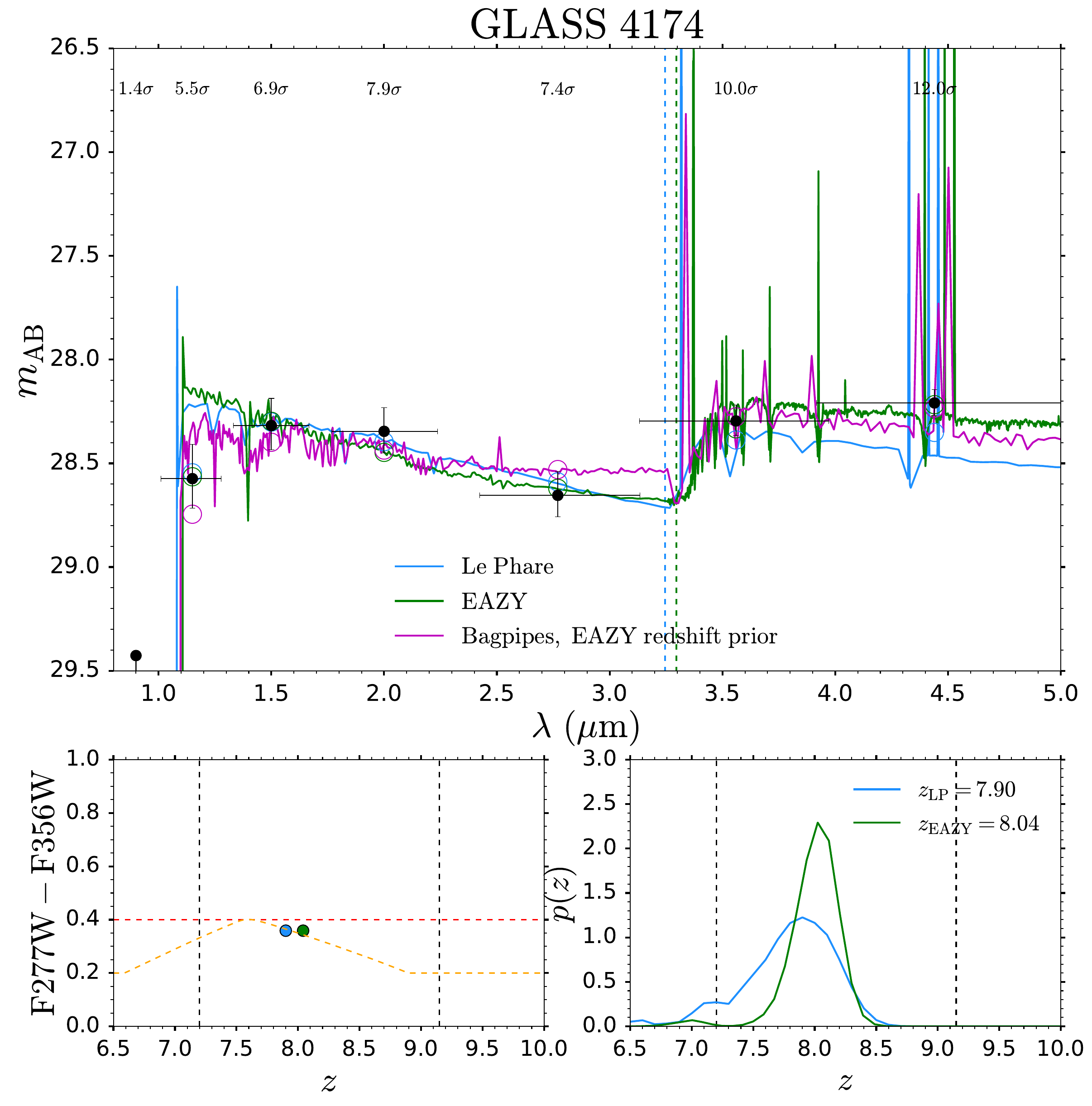} \hfill
\includegraphics[width=.475\linewidth]{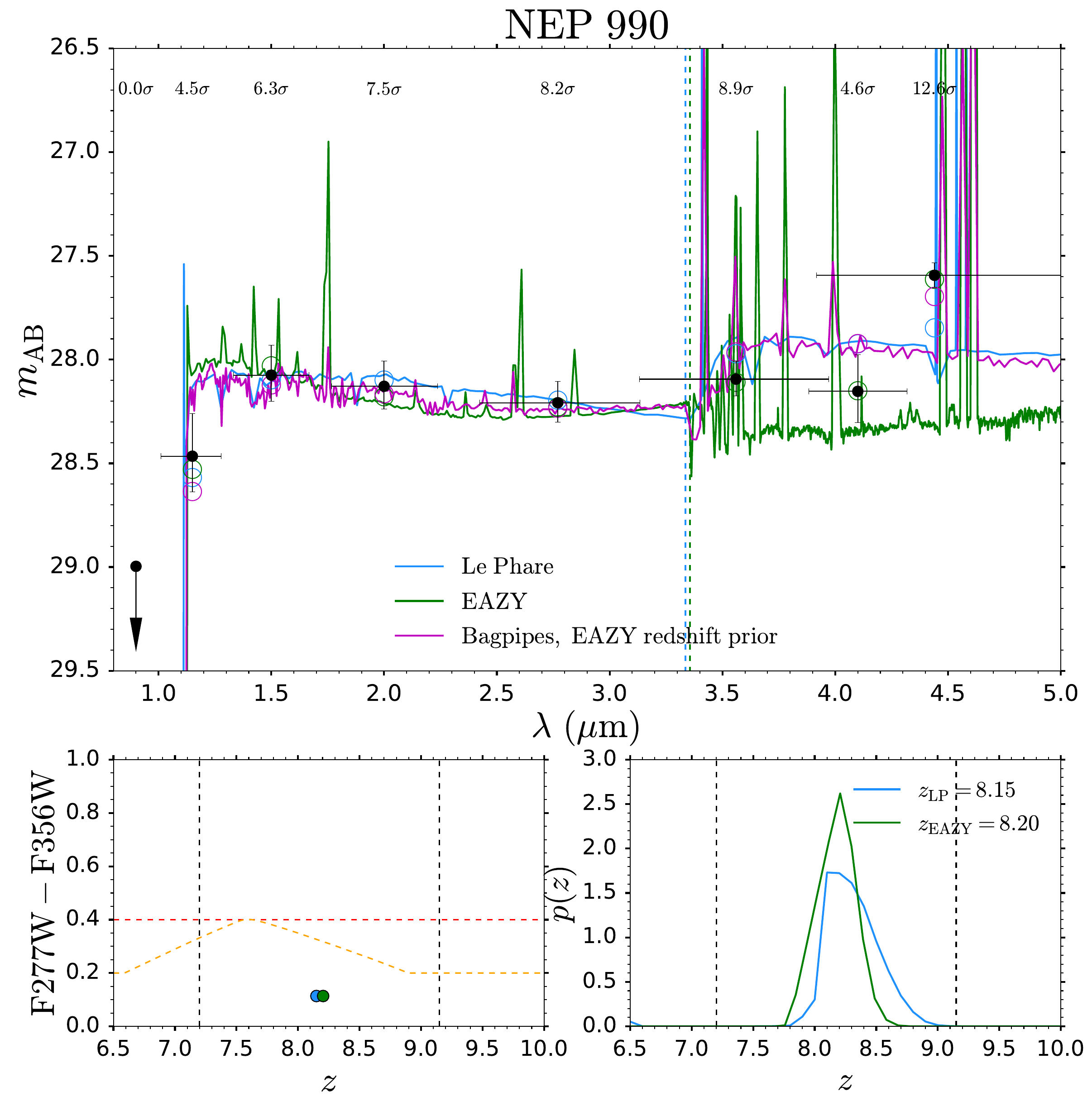} \\
\caption{Similar to Fig.~\ref{fig:f444w_candidates}, but now for our $z \sim 8$ sample.} 
\label{fig:f356w_candidates}
\end{figure*}

We show example SEDs for strict, relaxed and control galaxies in the top (left and right), bottom-left and bottom-right panels of Fig.~\ref{fig:f444w_candidates} (for our $z \sim 10.5$ sample) and Fig.~\ref{fig:f356w_candidates} (for our $z \sim 8$ sample), respectively. For each object, we display the NIRCam photometry in the top subpanel (in black), together with the best-fit Le Phare (light blue) and EAZY SEDs (green), as well as the median-fit Bagpipes SED (magenta, which has been fit assuming either the EAZY redshift PDF as a prior, or fixed at the known spectroscopic redshift). In the bottom subpanels, we show the position of these galaxies in our colour--redshift selection plane, as well as the Le Phare and EAZY redshift PDFs.

Our $z \sim 10.5$ sample, tabulated in Table~\ref{tab:f444w_candidates}, consists of 8 (33.3\%) strict Balmer-break candidates, 4 (16.7\%) relaxed Balmer-break candidates and 12 (50\%) control galaxies. Our $z \sim 8$ sample, tabulated in Table~\ref{tab:f356w_candidates}, consists of 4 (6\%) strict Balmer-break candidates, 7 (10\%) relaxed Balmer-break candidates and 57 (84\%) control galaxies. 

\begin{table*}
\begin{center}
\begin{tabular}{|c|c|c|c|c|c|c|c|c|c|} 
\hline
ID & Field & R.\@A.\@ & Dec.\@ & $z_\mathrm{phot}$ & F356W$-$F444W & $\log M_* (\mathrm{M}_\odot) $ & SFR ($\mathrm{M}_\odot~\mathrm{yr}^{-1}$)& Age$_\mathrm{MW}$ (Myr) & Sample \\
\hline
14109 & NGDEEP & 53.26715 & 27.84908 & $11.59^{+0.19}_{-0.20}$ & $0.49 \pm 0.12$ & $8.81^{+0.18}_{-0.37}$ & $4.19^{+1.77}_{-1.06}$ & $113^{+58}_{-81}$ & Strict \\
5416 & CEERS & 214.98596 & 52.87929 & $11.54^{+0.26}_{-0.35}$ & $0.50 \pm 0.30$ & $8.56^{+0.22}_{-0.39}$ & $2.55^{+1.39}_{-0.52}$ & $92^{+73}_{-62}$ & Strict \\
1998 & NEP & 260.71573 & 65.78262 & $11.09^{+0.26}_{-0.38}$ & $0.46 \pm 0.28$ & $8.61^{+0.17}_{-0.32}$ & $2.47^{+1.04}_{-0.50}$ & $111^{+68}_{-72}$ & Strict \\
4358 & NGDEEP & 53.24223 & 27.83051 & $10.84^{+0.26}_{-0.57}$ & $0.58 \pm 0.38$ & $8.48^{+0.25}_{-0.41}$ & $2.21^{+0.98}_{-0.50}$ & $88^{+92}_{-62}$ & Strict \\
1169 & CEERS & 214.95192 & 52.97168 & $10.71^{+0.14}_{-0.11}$ & $0.62 \pm 0.11$ & $9.90^{+0.15}_{-0.21}$ & $41.29^{+20.13}_{-12.12}$ & $142^{+52}_{-78}$ & Strict \\
6363 & NEP & 260.71332 & 65.7239 & $10.68^{+0.33}_{-0.17}$ & $0.48 \pm 0.19$ & $9.00^{+0.21}_{-0.24}$ & $6.17^{+3.12}_{-2.03}$ & $123^{+65}_{-68}$ & Strict \\
3953 & NEP & 260.82301 & 65.83401 & $10.62^{+0.14}_{-0.12}$ & $0.88 \pm 0.11$ & $9.55^{+0.12}_{-0.13}$ & $14.58^{+4.94}_{-3.37}$ & $176^{+31}_{-57}$ & Strict \\
2580 & CEERS & 214.81713 & 52.74835 & $10.30^{+0.16}_{-0.17}$ & $0.42 \pm 0.08$ & $9.48^{+0.10}_{-0.11}$ & $13.31^{+3.92}_{-2.67}$ & $160^{+47}_{-52}$ & Strict \\
\hline
545 & CEERS & 214.90665 & 52.94552 & $11.51^{+0.22}_{-0.17}$ & $0.23 \pm 0.14$ & $9.11^{+0.17}_{-0.27}$ & $7.94^{+3.39}_{-2.19}$ & $121^{+54}_{-77}$ & Relaxed \\
2401 & CEERS & 214.88724 & 52.86207 & $11.47^{+0.18}_{-0.24}$ & $0.37 \pm 0.35$ & $8.30^{+0.29}_{-0.38}$ & $2.31^{+0.70}_{-0.45}$ & $54^{+72}_{-37}$ & Relaxed \\
1915 & CEERS & 214.87584 & 52.91357 & $11.39^{+0.22}_{-0.27}$ & $0.33 \pm 0.20$ & $9.00^{+0.19}_{-0.29}$ & $6.66^{2.95}_{-1.95}$ & $110^{+62}_{-68}$ & Relaxed \\
1453 & CEERS & 215.01135 & 53.01177 & $11.34^{+0.17}_{-0.17}$ & $0.31 \pm 0.20$ & $8.58^{+0.27}_{-0.39}$ & $4.05^{+1.19}_{-0.74}$ & $61^{+70}_{-44}$ & Relaxed \\
\hline
\end{tabular}
\caption{Coordinates (Field, Right Ascension, Declination), EAZY photometric redshifts, F444W excess (F356W$-$F444W) and inferred physical properties (stellar mass $M_*$, SFR and mass-weighted stellar age) for our $z\sim10.5$ sample of strict and relaxed Balmer-break candidates. Note that the physical properties were derived using Bagpipes, with the masses and SFRs being corrected by the ratio of the fluxes measured within a Kron elliptical and 0.32~arcsec diameter circular aperture (see Section~\ref{subsec:mags_masses}).}
\label{tab:f444w_candidates}
\end{center}
\end{table*}

\begin{table*}
\begin{center}
\begin{tabular}{ |c|c|c|c|c|c|c|c|c|c|} 
\hline
ID & Field & R.\@A.\@ & Dec.\@ & $z_\mathrm{phot}$ & F277W$-$F356W & $\log M_* (\mathrm{M}_\odot) $ & SFR ($\mathrm{M}_\odot~\mathrm{yr}^{-1}$)& Age$_\mathrm{MW}$ (Myr) & Sample \\
\hline
4277 & NEP & 260.69567 & 65.72692 & $7.92^{+0.20}_{-0.20}$ & $0.44 \pm 0.21$ & $8.51^{+0.18}_{-0.24}$ & $1.89^{+0.59}_{-0.31}$ & $109^{+92}_{-59}$ & Strict \\
7508 & GLASS & 3.50867 & $-30.36662$ & $7.79^{+0.19}_{-0.19}$ & $0.73 \pm 0.29$ & $8.16^{+0.20}_{-0.28}$ & $1.47^{+0.44}_{-0.30}$ & $61^{+59}_{-34}$ & Strict \\
952 & NEP & 260.71194 & 65.79592 & $7.39^{+0.05}_{-0.26}$ & $0.46 \pm 0.12$ & $8.67^{+0.15}_{-0.15}$ & $7.53^{+2.15}_{-1.83}$ & $38^{+28}_{-16}$ & Strict \\
1456 & NEP & 260.76516 & 65.83038 & $7.21^{+0.17}_{-0.11}$ & $0.50 \pm 0.18$ & $8.50^{+0.17}_{-0.23}$ & $2.44^{+1.07}_{-0.84}$ & $84^{+100}_{-50}$ & Strict \\
\hline
660 & CEERS & 214.77137 & 52.84985 & $9.13^{+0.08}_{-0.20}$ & $0.27 \pm 0.23$ & $7.97^{+0.22}_{-0.28}$ & $3.22^{+0.86}_{-0.63}$ & $16^{+19}_{-10}$ & Relaxed \\
4858 & NEP & 260.74925 & 65.84637 & $9.08^{+0.08}_{-0.08}$ & $0.31 \pm 0.04$ & $9.50^{+0.10}_{-0.11}$ & $12.51^{+3.03}_{-1.81}$ & $174^{+59}_{-57}$ & Relaxed \\
12349 & NGDEEP & 53.25345 & $-27.79971$ & $8.95^{+0.12}_{-0.15}$ & $0.30 \pm 0.28$ & $8.13^{+0.18}_{-0.25}$ & $0.81^{+0.26}_{-0.15}$ & $110^{+81}_{-61}$ & Relaxed \\
6564 & CEERS & 214.77413 & 52.7385 & $8.86^{+0.12}_{-0.27}$ & $0.38 \pm 0.29$ & $9.25^{+0.14}_{-0.23}$ & $7.88^{+2.98}_{-1.65}$ & $160^{+80}_{-87}$ & Relaxed \\
3756 & NGDEEP & 53.23465 & $-27.81601$ & $8.53^{+0.20}_{-0.25}$ & $0.32 \pm 0.34$ & $8.40^{+0.18}_{-0.24}$ & $1.03^{+0.43}_{-0.22}$ & $175^{+75}_{-92}$ & Relaxed \\
6838 & NEP & 260.78402 & 65.85052 & $8.21^{+0.10}_{-0.12}$ & $0.37 \pm 0.06$ & $9.46^{+0.05}_{-0.05}$ & $7.26^{+1.32}_{-1.15}$ & $282^{+25}_{-44}$ & Relaxed \\
4174 & GLASS & 3.52019 & -30.36872 & $8.04^{+0.14}_{-0.25}$ & $0.36 \pm 0.18$ & $8.40^{+0.16}_{-0.20}$ & $1.61^{+0.50}_{-0.26}$ & $96^{+74}_{-44}$ & Relaxed \\
\hline
\end{tabular}
\caption{Similar to Table~\ref{tab:f444w_candidates}, but now for our $z\sim 8$ sample, also tabulating the F356W excess (F277W$-$F356W) rather than the F444W excess.}
\label{tab:f356w_candidates}
\end{center}
\end{table*}

\begin{figure}
\centering
\includegraphics[width=\linewidth]{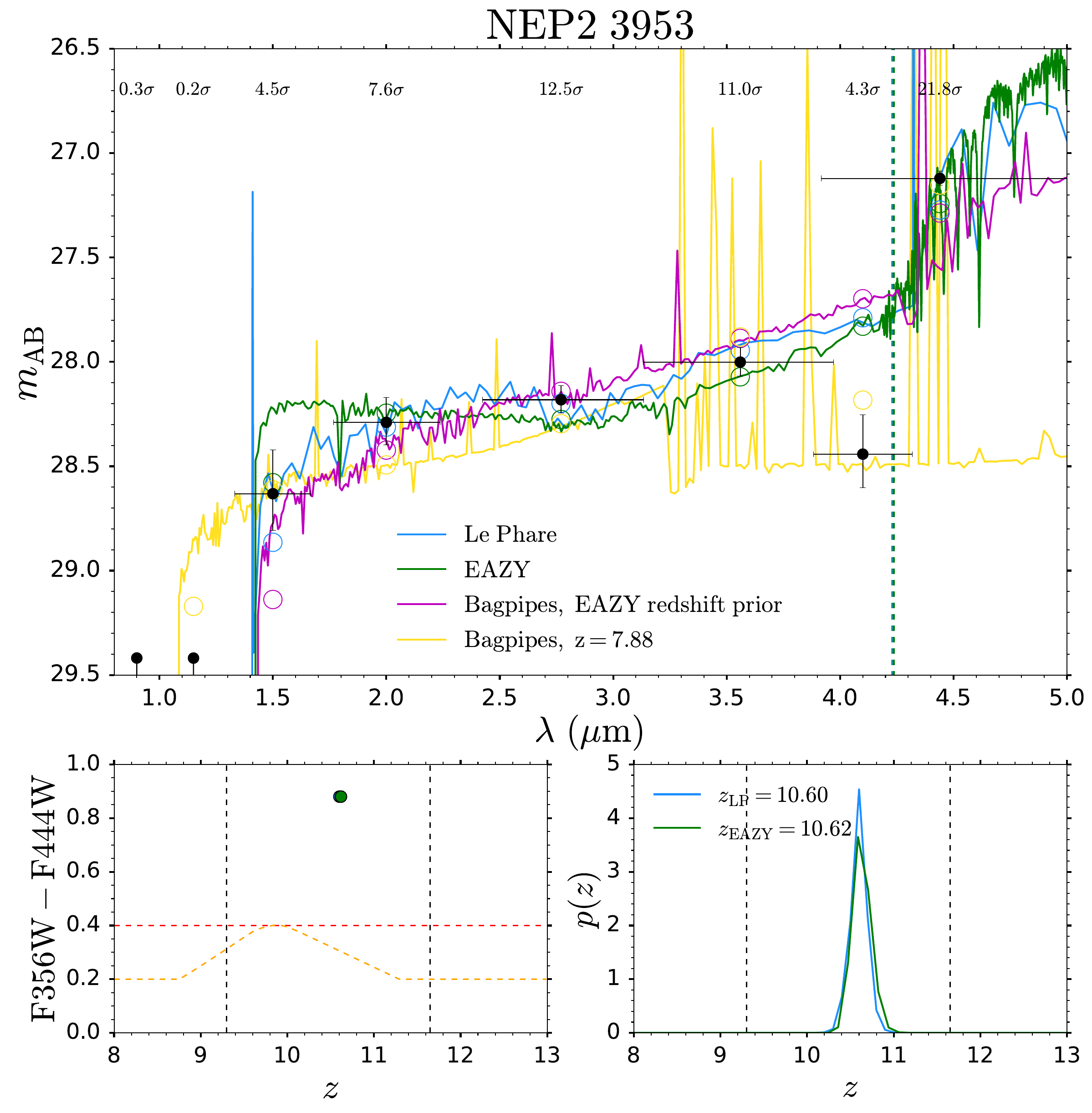}
\caption{The $z\sim10.5$ Balmer-break candidate displaying the largest F444W excess in our sample. Whilst the Balmer break solution (blue, green, purple) yields a satisfactory fit to the NIRCam wide-band data, the F410M flux density is strongly overpredicted. A $z=7.88$ Bagpipes fit (yellow) better matches the medium-band measurement, attributing the lack of F410M flux density to the low-lying Balmer jump continuum, with the F444W excess now being driven by strong \OIII\  + \Hb\ line emission. Additional medium-band photometry is needed to definitively distinguish between these two scenarios (see also Fig.~\ref{fig:flares_mediumband}).}
\label{fig:largest_break}
\end{figure}

We show the photometry and best-fitting SEDs for the $z\sim10.5$ Balmer-break candidate displaying the largest F444W excess in our sample in Fig.~\ref{fig:largest_break}. As can be seen from Fig.~\ref{fig:largest_break}, our standard model fits (blue, green, purple) provide a reasonable match to the wide-band NIRCam photometry. However, the exception lies with the medium-band F410M filter, whose bandpass-averaged flux density is strongly overpredicted by the models. Owing to the relatively low SNR measurement in this band ($4.3\sigma$), this may perhaps be due to noise, though this is still somewhat unlikely, as the F410M measurement deviates from the model-predicted value by $>$3$\sigma$, which has a likelihood of only $\sim$0.1\% of occurring.

The alternative explanation (missed by our Le Phare and EAZY SED fitting procedure) is that the F444W excess is attributable to strong \OIII\ + \Hb\ line emission, with the faint F410M flux density tracing the lowered rest-frame optical continuum level associated with the Balmer jump (essentially the inverse of the Balmer break, being driven by ionised hydrogen recombining to the $n=2$ state). This scenario requires the \OIII\ + \Hb\ lines to reside within the spectral range of the F444W filter, constraining the redshift to $7.0 < z < 9.3$. Furthermore, due to the dense packing of rest-frame optical emission lines, the F410M filter can only provide a clean (i.e.\@ unaffected by emission lines) measurement of the continuum when it falls between the \Hb\ and \Hg\ emission lines, which further constrains the redshift to $7.85 < z < 7.91$.  Hence we perform another Bagpipes fit (shown in yellow), with the redshift held fixed at $7.88$. We see from Fig.~\ref{fig:largest_break} that this yields a satisfactory fit to the observed data, with notably better agreement with the (somewhat noisy) F410M measurement.

As will be advocated throughout this paper, additional medium-band photometry is essential to reliably distinguish between photometric excesses driven by Balmer breaks and strong line emission, and hence is needed to definitively establish the nature of the source in Fig.~\ref{fig:largest_break}. Thus strong photometric excesses should be interpreted with particular caution in high-redshift galaxies with only wide-band photometry.  Such a system may be exhibiting a very prominent Balmer break and thus harbouring particularly old stellar populations, thus (in principle) placing valuable constraints on the onset of cosmic star formation. At the same time, the 0.88~mag excess in Fig.~\ref{fig:largest_break} demands a stellar population age of $>$150~Myr (see Fig.~\ref{fig:balmer_break_sb_cont}), which is a substantial fraction of the age of the Universe at that epoch ($\sim$450~Myr) and thus perhaps unlikely. On the other hand, strong line emission is ubiquitous in the EoR, with the 0.88~mag excess requiring a combined \OIII+\Hb\ rest-frame equivalent width of $\sim$1400~\AA, a condition readily met in these high-redshift systems \citep[e.g.\@][]{Trussler2023}.

\section{Results} \label{sec:results}

In this section we aim to establish how effective our Balmer break colour selections are at preferentially identifying galaxies with evolved stellar populations. We investigate and compare the physical properties of our Balmer-break candidates to those of the control sample, to establish whether there are any inherent differences between these two potentially distinct populations of galaxies. Additionally, we also discuss the qualitative constraints the inferred star formation histories for these Balmer-break candidates place on the onset of star formation in the Universe. 

We note that when we refer to `old' stellar populations or `old' stellar ages in this analysis, this is with respect to the age of the Universe at the epoch of observation. Since the Universe is still very young at these high redshifts (e.g.\@ 440~Myr at $z=10.5$), we consider galaxies with mass-weighted stellar ages of $\sim$100~Myr to be old in the following discussion. Of course, such galaxy ages would be considered to be very young at e.g.\@ $z=0$, where ages of several~Gyr are the norm \citep[see e.g.\@][]{Gallazzi2005, Gallazzi2021, Trussler2020, Trussler2021}.

\subsection{Magnitudes and stellar masses}\label{subsec:mags_masses}

\begin{figure*}
\centering
\includegraphics[width=\linewidth]{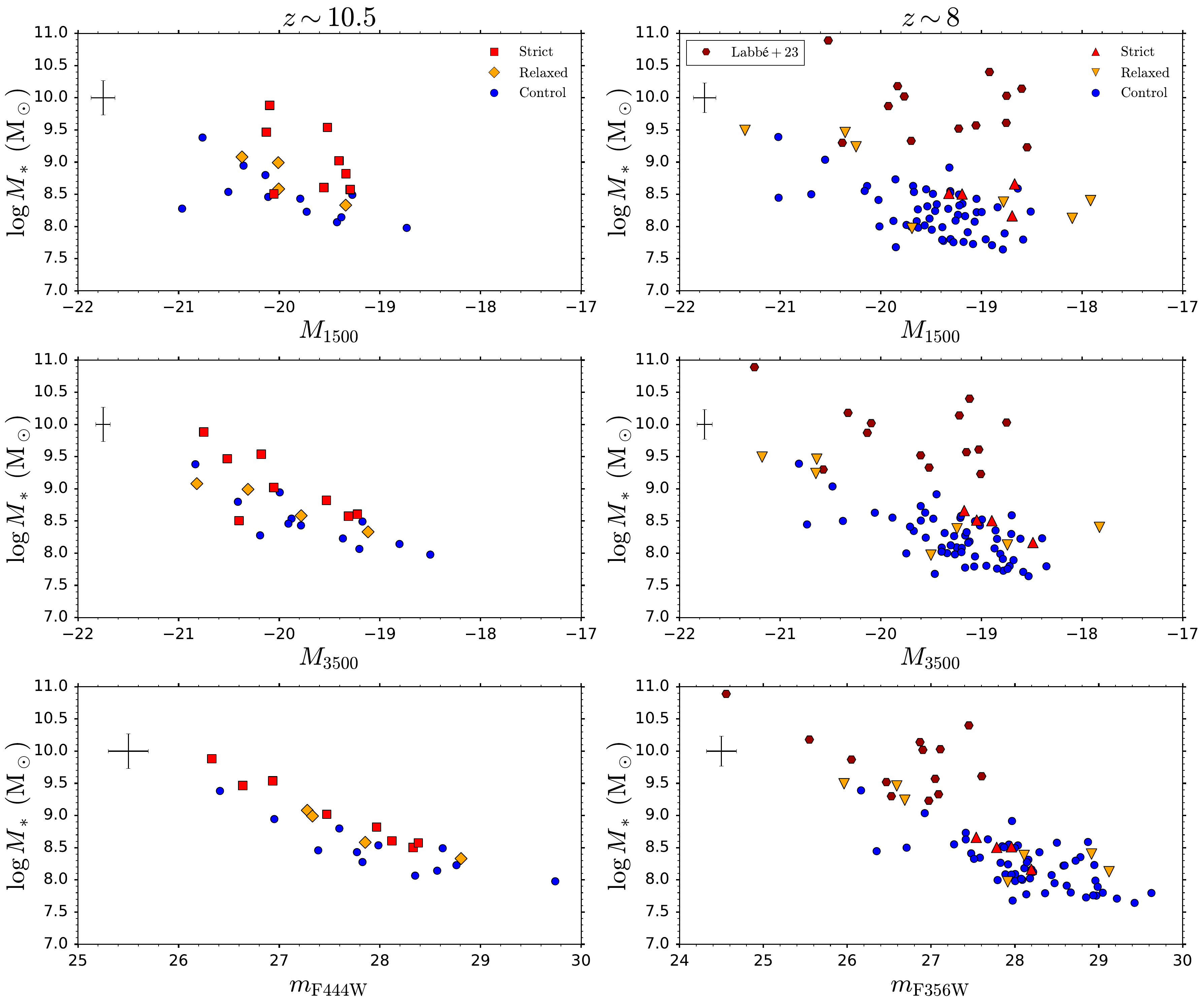}
\caption{Magnitudes (i.e.\@ brightnesses) against stellar masses for our $z \sim 10.5$ (left panels) and $z \sim 8$ samples (right), as well as the \citet{Labbe2023} sample of red massive candidate galaxies at $7.4 < z < 9.1$ (dark red hexagons). The median error bars for our samples are shown in the corner of each panel in black. Top panels: Rest-frame UV absolute magnitudes $M_{1500}$. Middle panels: Absolute magnitudes immediately blueward of the Balmer break ($M_{3500}$). Bottom panels: Rest-frame optical (i.e.\@ redward of the Balmer break) apparent magnitudes, using F444W (left) and F356W (right). Our Balmer-break candidates exhibit comparable intrinsic brightnesses to the control sample, but do tend to have larger inferred stellar masses for a given brightness. Thus the larger stellar masses found for our Balmer-break candidates are an outcome of the SED-fitting process (and the assumptions therein), rather than being due to their inherent brightness. As strong emission lines, the cumulative effect of weaker emission lines, a dusty continuum and a (dusty) AGN can all contribute to the red excess seen in these Balmer-break candidates, there is certainly room for bringing the (sometimes extremely large) inferred stellar masses in \citet{Labbe2023} down to be more in line with the expectations from simulations of galaxy formation.}
\label{fig:magnitudes}
\end{figure*}

We begin by comparing the magnitudes (i.e.\@ brightnesses) and stellar masses of our Balmer-break candidates against the control sample to determine whether there are any notable differences between the two, as well as to better understand the origin of the potentially (very) large stellar masses ($\log\, (M_*/\mathrm{M}_\odot) > 10$) inferred for some red rest-frame optical galaxies in the EoR \cite[see e.g.\@][]{Labbe2023}. 

We show the rest-frame UV (at 1500~\AA) absolute magnitudes (top panels), absolute magnitudes immediately blueward of the supposed Balmer break $M_{3500}$ (middle) and apparent magnitudes in the rest-frame optical (i.e.\@ redward of the Balmer break, bottom) against the Bagpipes-derived stellar masses in Fig.~\ref{fig:magnitudes}, with the $z \sim 10.5$ and $z \sim 8$ samples shown in the left and right panels, respectively. We have chosen to use apparent magnitudes rather than absolute magnitudes for the rest-frame optical, as we found that it can be difficult to accurately assess the continuum level redward of the Balmer break given our available data. As will be discussed later, this difficulty arises for the $z \sim 10.5$ sample simply due to the lack of photometry redward of the break. However, even for the $z \sim 8$ sample, with the F410M and F444W filters now probing beyond the Balmer break, there can be a degeneracy, with Balmer breaks, Balmer breaks plus line emission, and pure line emission sometimes being comparably compatible with the data. Hence a conclusive assessment of the continuum level cannot be made, as we cannot satisfactorily discriminate between these different scenarios without additional medium-band NIRCam and longer wavelength MIRI photometry. Note that we show the `total' magnitudes and `total' stellar masses, obtained by converting the $M_{1500}$ and $M_{3500}$ aperture magnitudes and stellar masses derived in the SED fitting process to `total' (i.e.\@ AUTO) magnitudes/masses. For this, we scale the inferred luminosity/mass by the ratio of the fluxes measured by the Kron elliptical aperture and the 0.32~arcsec diameter circular aperture in the appropriate filter. Namely, the F200W (F150W), F356W (F277W) and F444W (F356W) filters, for $M_{1500}$, $M_{3500}$ and the stellar mass, respectively, for the $z \sim 10.5$ ($z \sim 8$) sample.

From an inspection of Fig.~\ref{fig:magnitudes}, the strict + relaxed Balmer-break candidates and the control sample appear to be comparably distributed in terms of rest-frame UV magnitudes, as well as magnitudes blueward and redward of the Balmer break. Thus our Balmer-break candidates likely have comparable intrinsic brightnesses to the overall galaxy population at these redshifts, rather than being e.g.\@ systematically brighter. For reference, we also show the magnitudes and stellar masses (black hexagons) for the $7.4 < z < 9.1$ candidate massive galaxies from \citet{Labbe2023}, in the right panel of Fig.~\ref{fig:magnitudes}. Whilst these objects are comparably bright to our full $z \sim 8$ sample in the rest-frame UV (as traced by \@ $M_{1500}$), they do tend to be slightly brighter immediately blueward of the Balmer break (as traced by $M_{3500}$), and much brighter redward of the Balmer break (as traced by $m_\mathrm{F356W}$). This is not too surprising, given the \citet{Labbe2023} criteria for selecting these massive candidates. Namely, bright ($m_\mathrm{F444W} < 27$~AB mag), very red (F277W$-$F444W > 1) galaxies. 

Now, what we do find for our own sample, is that the Balmer-break candidates do tend to have larger stellar masses than the control sample for a given brightness. Thus the larger stellar masses found for our Balmer-break candidates are an outcome of the SED-fitting process, rather than being due to their inherent brightness. As we only include the light contributions from stellar populations and nebular emission in our Bagpipes fitting process (i.e.\@ excluding an AGN component),  these larger stellar masses are attributable to requiring an older stellar population component, with higher mass-to-light ratios, to best reproduce the data. Although the \citet{Labbe2023} sample consists of brighter galaxies than our sample, it is still possible that the large stellar masses inferred for their systems (shown in Fig.~\ref{fig:magnitudes}) are perhaps also still an outcome of the SED-fitting process, rather than being due to (and thus demanded by) the inherent brightness of these galaxies.

\begin{figure*}
\centering
\includegraphics[width=\linewidth]{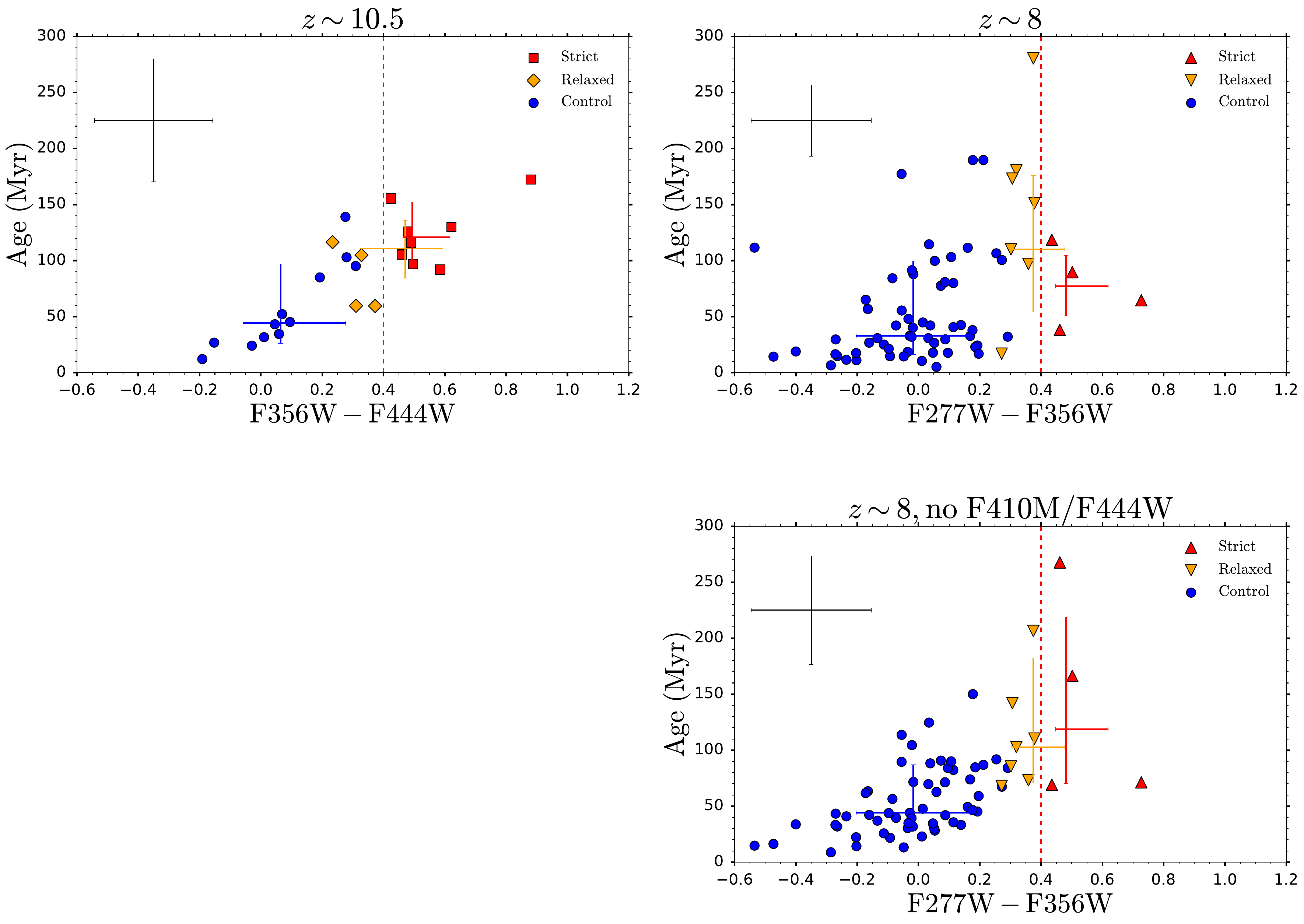}
\caption{Photometric excess (a proxy for the Balmer break strength) against the inferred mass-weighted stellar ages of the galaxies in our sample. Colour coding and symbols are the same as in Fig.~\ref{fig:balmer_break_selection}. The median excess and age are also shown for each sample, with the endpoints of the error bars representing the 16--84 percentile range. The median and percentile range for the relaxed sample combines both the relaxed and strict sample galaxies. The red dashed vertical line denotes our flat colour threshold for selecting strict Balmer-break candidates. Top-left panel: F444W excess for our $z \sim 10.5$ sample of galaxies. Top-right panel: F356W excess for our $z \sim 8$ sample of galaxies. Note the weaker correlation between the observed excess and the inferred stellar age than for the $z \sim 10.5$ sample. Bottom-right panel: F356W excess for our $z \sim 8$ sample of galaxies, but now omitting the F410M and F444W photometry in the SED-fitting with Bagpipes (mimicking the rest-frame coverage for the $z \sim 10.5$ sample). A stronger correlation between observed excess and age emerges, as with the $z \sim 10.5$ sample. Thus, longer wavelength photometry, which probes the (stronger) emission lines in the rest-frame optical beyond the Balmer break, are likely essential to derive more accurate stellar ages and star formation histories for galaxies (such as MIRI F560W for our $z \sim 10.5$ sample).}
\label{fig:colour_age}
\end{figure*}

As we will discuss in more detail in Section~\ref{sec:discussion}, strong emission lines (i.e.\@ \OIII\ and \Hb), the cumulative effect of weaker emission lines, a dusty continuum and a (dusty) AGN can all contribute to the red excess seen in the supposed Balmer-break candidates. All of these effects alleviate the need for a substantially old (${\sim}100$~Myr) stellar population, i.e.\@ reduce the inferred stellar mass-to-light ratio, thus bringing the resulting (possibly extremely large) stellar masses inferred in \citet{Labbe2023} down to be more in line with the expectations from simulations of galaxy formation \citep[see e.g.\@][]{Lovell2023}.

\subsection{Ages and star formation histories}

In order to establish how well our Balmer break colour selections are at identifying galaxies with evolved stellar populations, we show the measured F444W excess (i.e.\@ F356W$-$F444W for our $z\sim 10.5$ sample, top-left) and F356W excess (i.e.\@ F277W$-$F356W for our $z\sim 8$ sample, top- and bottom-right) against the inferred mass-weighted stellar ages (derived using Bagpipes, see Section~\ref{sec:data}) in Fig.~\ref{fig:colour_age}. As before, the strict, relaxed and control samples are shown in red, orange and blue, respectively. We also show the median excess colour and age for each sample, with the endpoints of the error bars representing the 16--84 percentile range. Note that the median and percentile range for the relaxed sample represent the median and percentile range for the combined strict and relaxed samples (rather than just the relaxed sample alone). 

From the left panel of Fig.~\ref{fig:colour_age}, we see that the observed F444W excess appears to be a good predictor of the mass-weighted stellar age of the $z \sim 10.5$ galaxies, with the inferred age generally increasing with increasing F444W excess. Indeed, our Balmer-break candidates have a median inferred age of 115~Myr versus the 50~Myr of the control sample. From this panel alone, this result suggests that the observed strength of (what we believe to be) a Balmer break places tight constraints on the age and star-formation history of a galaxy. We note that the four control galaxies with ages $>90$~Myr all narrowly miss our redshift-dependent colour cut.

\begin{figure*}
\centering
\includegraphics[width=\linewidth]{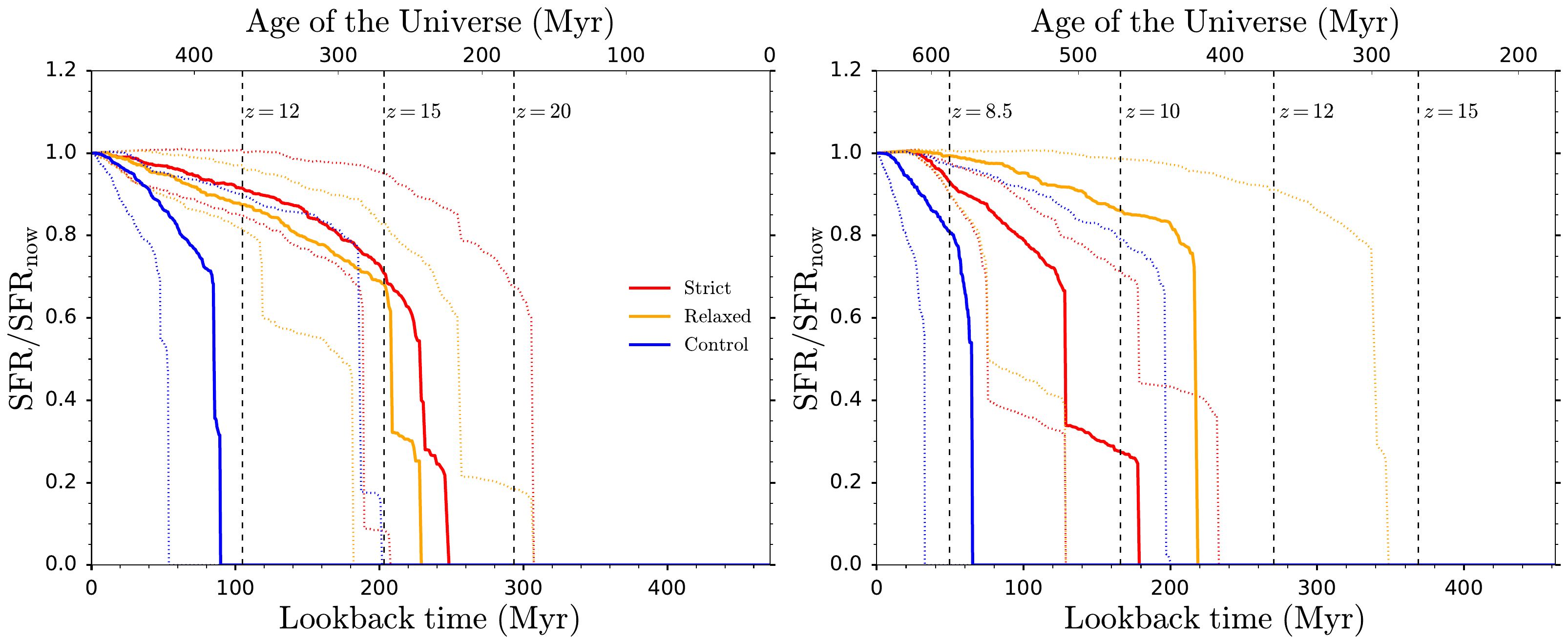}
\caption{Normalised (to the current SFR) star formation histories for our $z \sim 10.5$ (left panel) and $z \sim 8$ samples (right). We display both the median (solid) and 16, 84 percentiles (dotted) SFH percentiles, determined by taking the appropriate percentile per lookback time bin of the normalised SFH for each galaxy in the sample. The age of the Universe on the upper axis assumes the galaxy is at $z=10.5$ (left) and $z = 8$ (right). We stress that we just focus on the qualitative aspects of the inferred star formation histories, due to the insufficient constraining power from the current data, as well as the dependency on the assumed SFH parameterisation and priors on the associated parameters. Ultimately, deep continuum spectroscopy with the NIRSpec PRISM, together with MIRI imaging, will be needed to provide the best possible indirect constraints on the onset of star formation in the Universe.}
\label{fig:sfh}
\end{figure*}

However, from the top-right panel of Fig.~\ref{fig:colour_age}, where we now show the F356W excess for our $z \sim 8$ sample, we see that this potential connection between the (supposed) observed Balmer break strength and galaxy age is much less clear. Galaxies with a strong F356W excess (our strict sample) do not necessarily have old stellar ages. Additionally, galaxies with a weak F356W excess (our control sample) can still have old stellar ages. What is true, however, is that our flat colour selection (red) and redshift-dependent colour selection (orange) do indeed, on average, select older galaxies (70, 100~Myr vs.\@ the 30~Myr for the control sample). Though this trend is less strong than what was seen at $z \sim 10.5$.

The key difference, from the observational perspective, for our $z \sim 8$ sample is that, being at lower redshift, the F410M and F444W filters now probe at longer rest-frame wavelengths than for our $z \sim 10.5$ sample. More specifically, these filters now probe the rest-frame optical beyond the Balmer break (as F356W probes the break itself), crucially covering the rest-frame optical emission lines (such as \OIII\ and \Hb). As the strength (i.e.\@ equivalent width) of these emission lines is closely connected to recent star formation \citep[or shocks and AGN activity, see e.g.\@][]{Baldwin1981, Kewley2019}, the F410M and F444W filters thus provide valuable additional constraints on the current star formation rates in these galaxies, thus likely refining the inferred star formation histories and stellar ages (though we do note that there can still be degeneracies, with Balmer breaks, breaks plus line emission and pure line emission sometimes being comparably compatible with the data).

To establish whether this is indeed the case, or whether the different results at $z \sim 10.5$ and $z \sim 8$ are driven by a fundamental difference between the star formation histories of galaxies at these two redshifts, we repeat our Bagpipes SED-fitting procedure for our $z \sim 8$ sample. We now intentionally omit the F410M and F444W photometry, with F356W (probing the Balmer break) thus being the longest available filter, hence mimicking the rest-frame coverage for our $z \sim 10.5$ sample (with F444W probing the Balmer break). We show the results of this revised SED-fitting procedure in the bottom-right panel of Fig.~\ref{fig:colour_age}. We find, similar to our $z \sim 10.5$ analysis, that the excess correlates more strongly with the inferred stellar age. This is in contrast to the weaker correlation seen in the middle panel where the added constraints from F410M and F444W are included.

Hence longer wavelength photometry, which probes the (stronger) emission lines in the rest-frame optical beyond the Balmer break, are likely essential to derive more accurate stellar ages and star formation histories for galaxies. Thus, for the highest redshift galaxies (such as our $z \sim 10.5$ sample), additional MIRI imaging (e.g.\@ F560W) will be crucial to place the tightest possible indirect constraints on the onset of star formation in the Universe.

Now, the weakened connection between the (supposed) observed Balmer break strength (i.e.\@ F356W excess) and the inferred galaxy stellar age seen in the top-right panel of Fig.~\ref{fig:colour_age} should not necessarily be surprising. As has already been known for dwarf galaxies in the local Universe \citep[see e.g.][]{Emami2019}, and is now becoming more definitely established at high-redshift with \emph{JWST}, the star formation histories of galaxies in the epoch of reionisation are likely bursty \citep[see e.g.\@][]{Endsley2023, Looser2023, Looser2023b}. Indeed, galaxies in this era can likely display both elevated star formation activity and a (temporary) cessation \citep{Looser2023} of star formation. Thus there can be a disconnect between the current star formation activity (and hence key SED properties) and the extended star formation histories of these galaxies. As we will discuss in more detail in Section~\ref{sec:discussion}, it is for this reason that a (currently observed) Balmer break need not necessarily imply a particularly old galaxy \citep[see also][]{Wilkins2023}. Likewise, a galaxy harbouring evolved stars may simply have its older stars outshone \citep[see e.g.\@][]{Narayanan2023} by the brighter younger stars (thus resulting in a weak Balmer break, if any).

With the lack of MIRI F560W constraints for our $z \sim 10.5$ sample in mind, together with the possible degeneracy between line emission and breaks even for our $z \sim 8$ sample, we now turn to the inferred star formation histories for these high-redshift galaxies. In principle, the presence of evolved stars (inferred through their unique spectral signatures) in high-redshift galaxies observed shortly after the Big Bang places valuable constraints on the formation history of the first galaxies in the Universe. Indeed, taken at face value, the mass-weighted stellar ages of $\sim$150~Myr and $\sim$275~Myr at $z\sim 10.5$ ($\mathrm{age}_\mathrm{Universe} = 440$~Myr) and $z \sim 8$ ($\mathrm{age}_\mathrm{Universe} = 640$~Myr) correspond to redshifts of $z = 14.2$ and $z = 12$, respectively. In practice however, due to the aforementioned caveats, together with the fact that the star formation histories derived likely depend on the SFH parameterisation assumed in the SED-fitting process, as well as the priors placed on these parameters, strict caution should be applied in interpreting these results. 

Hence we just focus on the qualitative aspects of the inferred star formation histories, which we show in Fig.~\ref{fig:sfh}. For each sample, we determine the normalised (to the current SFR) star-formation history (inferred using the SED-fitting code Bagpipes), by taking the median (or 16, 84 percentile) per lookback time bin of the normalised SFH for each galaxy in the sample. In line with their older inferred mass-weighted stellar ages, our strict and relaxed Balmer-break candidates tend to have more extended star formation histories than the control sample. As will be discussed in more detail in Section~\ref{sec:discussion}, additional NIRCam medium-band imaging will be essential to better disentangle the effects of line emission and Balmer breaks, as well as to discriminate between evolved stellar populations and dusty galaxies/AGN. Ultimately, deep continuum spectroscopy with the NIRSpec $R$$\sim$100 PRISM (the NIRISS and NIRCam grisms lack the necessary continuum sensitivity), together with MIRI imaging, will be needed to provide the best possible indirect constraints on the onset of star formation in the Universe. 

\subsection{Further star formation properties}

\begin{figure*}
\centering
\includegraphics[width=\linewidth]{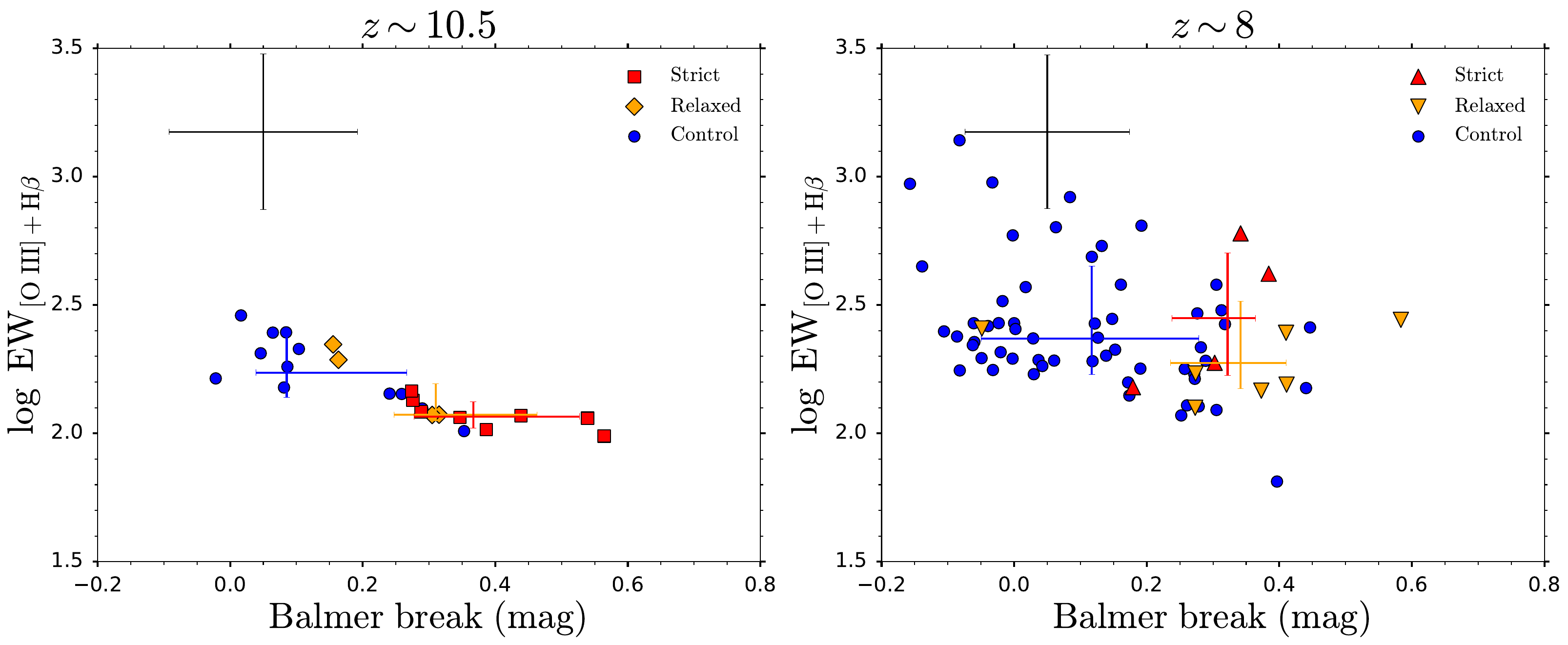}
\caption{Balmer break strength (in magnitudes, see text for definition) against the combined rest-frame equivalent width of the \OIII\ $\lambda\lambda 4959, 5007$ doublet and \Hb, both inferred from SED fitting. It should be noted that the \OIII\ and \Hb\ lines are redshifted out of the NIRCam range for our $z \sim 10.5$ sample, with the inferred equivalent widths therefore being based off an extrapolation with no direct/reliable emission line constraints in principle being available from the photometry. Our strict and relaxed colour selections do tend to select galaxies with larger Balmer breaks, though these galaxies can have comparable emission line equivalent widths to the control sample. Furthermore, galaxies that do not meet our colour selection criteria are still found to have large ($>$0.4~mag) Balmer breaks. Hence, ideally the full available photometry should be used to identify galaxies with strong Balmer breaks/evolved stellar populations (as opposed to the single colour cut in this work), though additional medium-band NIRCam imaging will be essential to reliably make such inferences from the photometric data (see Section~\ref{sec:discussion}).}
\label{fig:break_ew}
\end{figure*}

We now turn to further inferred SF-related properties for our $z \sim 10.5$ and $z \sim 8$ samples. Indeed, prior we examined the observational proxy for the Balmer break strength, namely the colours $C_{34} =$ F356W$-$F444W and $C_{23} =$ F277W$-$F356W. However, these measured colours are subject to observational errors, with the perceived colour also being dependent on the redshift of the source, and can also be affected by the flux from weak emission lines (which we will discuss in more detail in Section~\ref{sec:discussion}). Thus, here we instead focus on the Balmer break strength inferred from a full SED fit to the data, which should in principle account for these effects. 

In Fig.~\ref{fig:break_ew}, we show the inferred Balmer break strength (in magnitudes) against the combined rest-frame equivalent width of the \OIII\ $\lambda\lambda 4959, 5007$ doublet and \Hb\ (inferred from SED fitting). Here we define the Balmer break strength to be the ratio of the median flux density (in terms of $f_\nu$) between 4000--4300~\AA\ (rest-frame) and 3400--3600~\AA, i.e.\@ redward and blueward of the Balmer break. Thus our definition of the Balmer break strength spans both the Balmer break itself (at 3646~\AA\ rest-frame), as well as the well-known 4000~\AA\ break. We have chosen to use this broad wavelength range to mimic the rest-frame wavelength ranges typically probed by e.g.\@ the (F356W, F444W) or (F277W, F356W) filter pairs in our observational analysis. 

\begin{figure*}
\centering
\includegraphics[width=\linewidth]{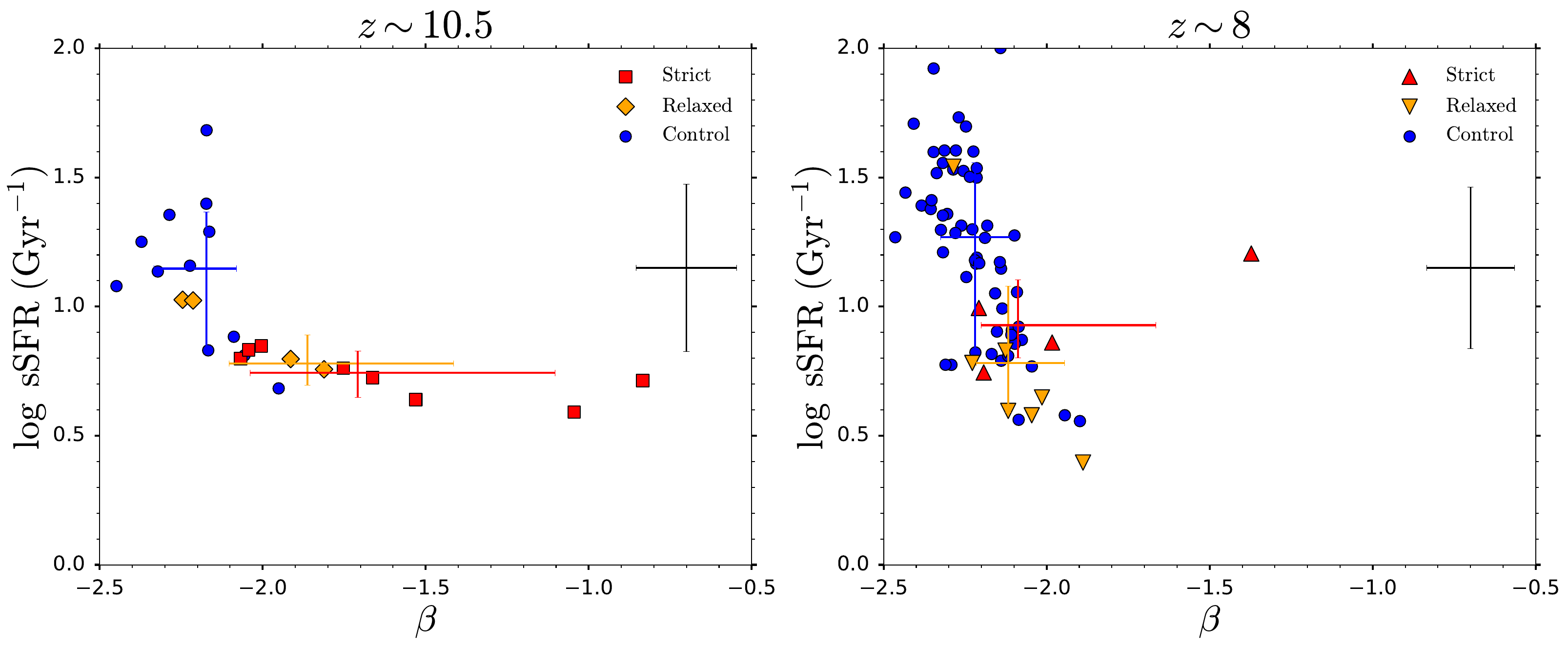}
\caption{UV slope $\beta$ against specific star formation rate (sSFR, using the average SFR over the past 10~Myr). Our strict and relaxed Balmer-break candidates at $z \sim 10.5$ tend to have lower specific star formation rates and redder $\beta$ slopes than the control sample, though the $\beta$ trends seem less strong at $z \sim 8$.}
\label{fig:beta_sSfr}
\end{figure*}

As can be seen from the left panel of Fig.~\ref{fig:break_ew}, our strict and relaxed Balmer break colour selection does tend to select galaxies with stronger inferred Balmer break strengths at $z \sim 10.5$ (with a median of 0.35, 0.32~mag vs.\@ 0.09~mag for the control sample). From the SED-fitting these galaxies are expected to have weaker (i.e.\@ lower equivalent width, with a 120~\AA\ vs.\@ a 160~\AA\ median) line emission though note that the \OIII\ and \Hb\ emission lines are (by design of our redshift cut) redshifted out of the F444W filter, so no direct/reliable emission line constraints are in principle available from the photometry. Indeed, at $z \sim 8$ (right panel), where F410M and F444W now do provide direct constraints on the emission line strengths, we find that our strict and relaxed Balmer-break candidates, together with the control sample, have comparable inferred line equivalent widths. On average, our strict and relaxed samples do tend to exhibit stronger (0.33, 0.35~mag vs.\@ 0.11~mag) inferred Balmer breaks (though again we stress that with the available photometry, there can still be degeneracy between the relative contributions from emission lines and the Balmer break). We note that there are also control galaxies which exhibit large Balmer breaks ($>$0.4~mag). These objects either narrowly miss our redshift-dependent colour cut, or are near the upper edge of our redshift cut, where the sensitivity (through the F277W$-$F356W colour) to the Balmer break drops as the break gets redshifted out of the F356W filter. 

\begin{figure*}
\centering
\includegraphics[width=\linewidth]{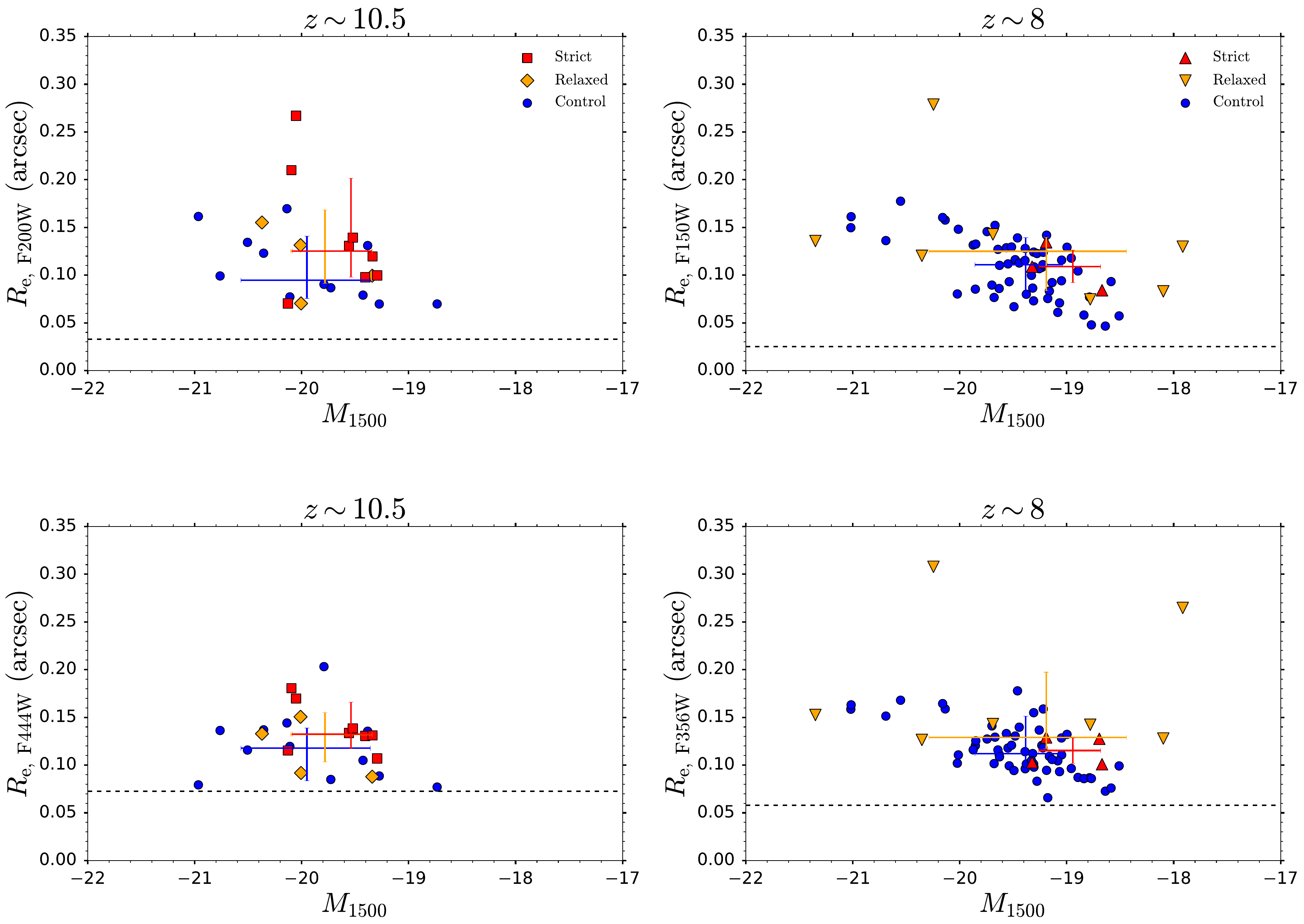}
\caption{Effective radius $R_\mathrm{e}$ (in arcsec) against absolute UV magnitude $M_{1500}$. Top panels: Rest-frame ultraviolet sizes (F200W and F150W for our $z \sim 10.5$ and $z \sim 8$ samples, respectively). Bottom panels: Rest-frame optical sizes (F444W and F356W for our $z \sim 10.5$ and $z \sim 8$ samples, respectively). We find that the strict+relaxed Balmer-break candidates are roughly comparably distributed in effective radii to the control galaxies, suggesting that the typical relative contribution of AGN emission (if any) is likely similar in these systems. Some of our Balmer-break candidates (as well as control galaxies), particularly in the rest-frame optical, have small effective radii that are only slightly above the filter half width half maximum (dashed horizontal lines). Thus it is plausible that these systems may possibly harbour an AGN component, which have been suggested to potentially contribute to the flux excess seen in the longer wavelength NIRCam bands.}
\label{fig:MUV_Re}
\end{figure*}

This therefore highlights the need to, ideally, use the full available photometry to identify galaxies with strong Balmer breaks and evolved stellar populations (as opposed to a single colour cut as in this work). However, as will be discussed in more detail in Section~\ref{sec:discussion}, additional medium-band NIRCam imaging will be essential to reliably make such inferences from the photometric data.

Furthermore, we note that, upon omitting the F410M and F444W photometry (as before) for the $z \sim 8$ sample and refitting with Bagpipes, we obtain a strong anticorrelation between Balmer break strength and the emission line equivalent width (as for the $z \sim 10.5$ sample, not shown).

As mentioned in Section~\ref{sec:intro}, the 4000~\AA\ break also serves as a stellar age indicator, typically building up over $\sim$Gyr timescales (compared to the $\sim$several hundred Myr timescales for the Balmer break). Hence the 4000~\AA\ break strength is expected to be small for the high-redshift, epoch of reionisation galaxies considered in this work. Indeed, we find that the median 4000~\AA\ break strength for our $z \sim 10.5$ sample of strict Balmer-break candidates is only 0.14~mag and 0.04~mag when using the original \citep{Bruzual1983} and narrow \citep{Balogh1999} definitions of the 4000~\AA\ break, respectively. These are considerably smaller than our $C > 0.4$~mag requirement for selecting strict Balmer-break candidates. Hence we advise against using the 4000~\AA\ break strength (alone) as an age indicator for EoR galaxies, due to it being so small (at these early epochs) and therefore difficult to measure both through spectroscopy and through photometry. Rather, we instead advocate for using the stronger Balmer break, or the combined Balmer + 4000~\AA\ break, as a stellar age indicator, though see Section~\ref{subsec:break_caveats} for caveats regarding this.

Finally, we briefly comment on the inferred specific star formation rates (sSFR, using the average SFR over the past 10~Myr) and UV slopes $\beta$ for these galaxies, which we show in Fig.~\ref{fig:beta_sSfr}. Again, although we find a clear connection between the sSFR and $\beta$ UV slope at $z \sim 10.5$, with our strict and relaxed Balmer-break candidates tending to have lower specific star formation rates (6~Gyr$^{-1}$ vs.\@ 12~Gyr$^{-1}$) and redder $\beta$ slopes ($-1.7$, $-1.9$ vs.\@ $-2.2$), we find that the $\beta$ trends are less strong at $z \sim 8$. 

\subsection{Galaxy sizes}

Here we comment on the apparent sizes (measured using the SExtractor FLUX\_RADIUS parameter) of our high-redshift galaxies, aiming to establish whether any of our Balmer-break candidates have compact morphologies compatible with a potentially dominant AGN component (thus disfavouring the Balmer break scenario). We show the rest-frame ultraviolet effective radii (probed by F200W for $z \sim 10.5$ and F150W for $z \sim 8$, top panels), as well as the rest-frame optical effective radii (probed by F444W and F356W respectively, bottom panels) in Fig.~\ref{fig:MUV_Re}. The horizontal dashed lines denote the half width half maximum of the filters. For both our $z \sim 10.5$ and $z \sim 8$ samples, we find that the strict+relaxed Balmer-break candidates are roughly comparably distributed in effective radii to the control galaxies, rather than being e.g.\@ preferentially more compact if a dominant AGN component was often present in these systems. Thus, the typical relative contribution of AGN emission (if any) is likely similar for the Balmer-break candidates and control galaxies. We do note that some of our Balmer-break candidates (as well as control galaxies), particularly in the rest-frame optical, have small effective radii that are only slightly above the filter half width half maximum. Thus it is plausible that these systems may potentially harbour an AGN component, which have been suggested \citep[see e.g.\@][]{Furtak2022, Barro2023, Kocevski2023, Labbe2023b, Matthee2023b} to contribute to the flux excess seen in the longer wavelength NIRCam bands for some red-looking galaxies in other samples \citep{Labbe2023}.

\section{Reliably identifying Balmer breaks and their constraining power} \label{sec:discussion}

In this section we discuss caveats regarding the identification of Balmer breaks from photometric data, as well as the constraints these Balmer breaks can place on the presence of evolved stellar populations in high-redshift galaxies.

\subsection{Strong line emission}

The basis behind the lower limit for our redshift cut is to ensure that the strong rest-frame optical lines \OIII\ and \Hb\ are both redshifted out of the filter of interest, thus increasing the likelihood that the photometric excess seen is attributable to a Balmer break. However as the bulk of our redshifts are based off of photometric redshifts, there can be some uncertainty, especially with the (limited) available photometry, that this is indeed the case. The two defining features in the SEDs of these high-redshift Balmer-break candidates are the Lyman break in the shorter wavelength bands, together with the (supposed) Balmer break excess seen in the F444W or F356W bands. 

However, while the perceived photometric strength and location of the Lyman break depend on the source redshift (and depth), galaxies can also begin to dropout in these same filters due to strong dust attenuation in the rest-frame UV. In order to reproduce the same observed dropout strength, dropouts that are driven purely by IGM Ly$\alpha$ attenuation will therefore be assigned a higher photometric redshift than galaxies that dropout by a combination of IGM and dust attenuation. If there is insufficient flexibility in the dust models used in the SED-fitting process to mimic the dust attenuation laws seen in these high-redshift galaxies, this could cause the redshift to be overestimated.

In addition, insufficient flexibility in the emission line templates can also cause the observed red F444W (or F356W) excess to be preferentially fit by a higher-redshift Balmer break, causing the lower-redshift \OIII + \Hb\ solution to be overlooked entirely (thus yielding close to zero probability in the resulting photometric redshift PDFs). We found that this could be the case with e.g.\@ the SED-fitting code Le Phare, where the emission line strengths (i.e.\@ equivalent widths) in the models are often too low to reproduce the observed data, thus heavily favouring the Balmer break solution \citep[see also][]{Adams2023, Trussler2023}.

\begin{figure}
\centering
\includegraphics[width=\linewidth]{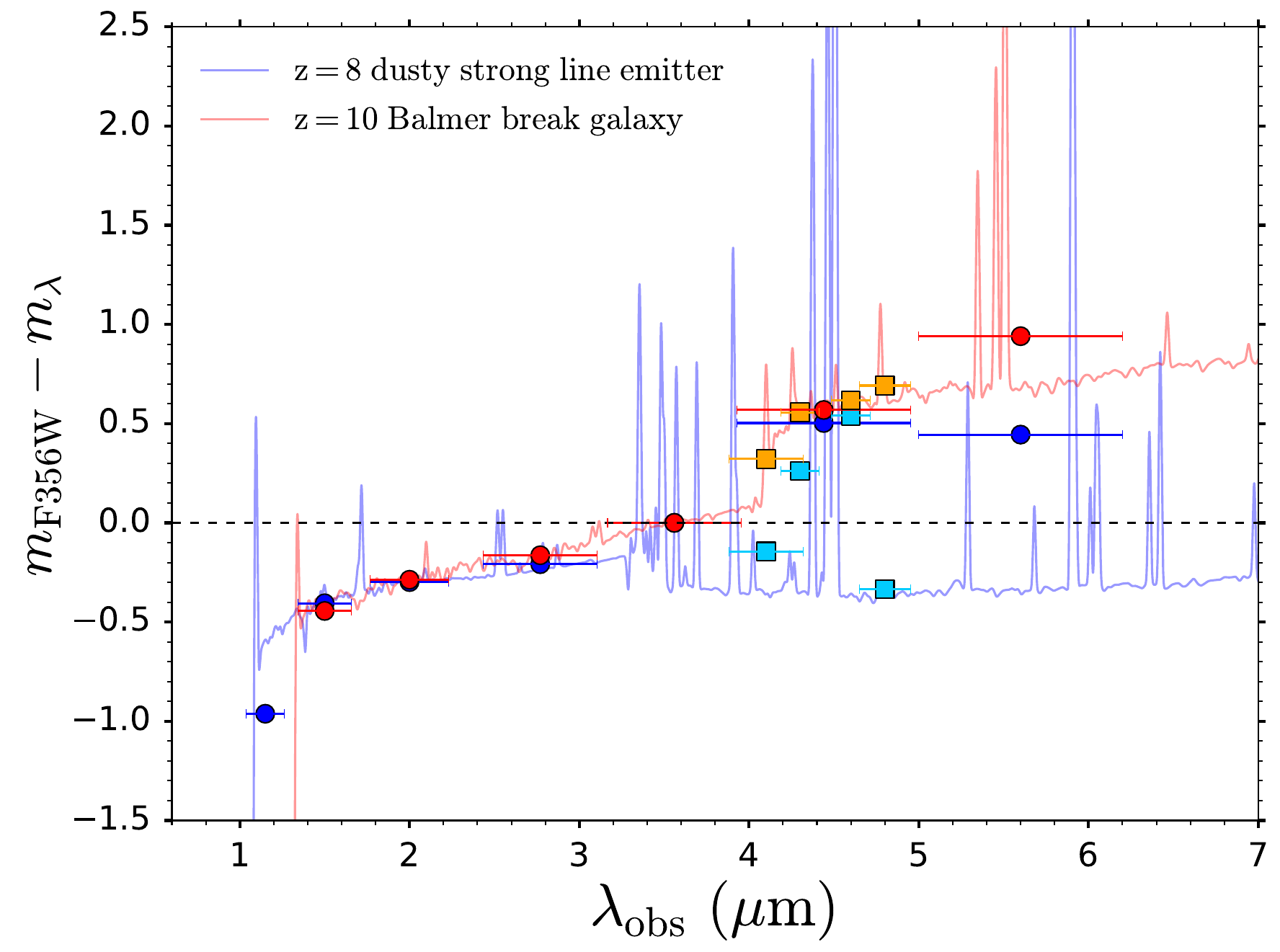}
\caption{SEDs and photometry for two galaxies from the FLARES simulations. The magnitude difference $m_{F356W} - m_{\lambda}$ on the vertical axis is the magnitude offset relative to F356W, to highlight the strength of the (supposed) Balmer break. We show both a $z=8$ strong line-emitting galaxy (blue) and a $z=10$ Balmer-break galaxy (red and orange). These two galaxies were selected to have very similar NIRCam wide-band photometry (dark blue and red circles, below 5~\textmu m), hence it would be difficult to discriminate between these two scenarios, given only this data. However, with the addition of NIRCam medium-band photometry (light blue and orange squares), one can readily distinguish the two, as the signature of the $z=10$ Balmer break is revealed through the consistently elevated flux density in the medium-band filters, while the medium-band photometry for the $z=8$ line emitter varies between being high or low depending on whether it is boosted by line emission or just tracing the continuum level. Furthermore, MIRI F560W imaging (dark blue and red circles, above 5~\textmu m) can in principle also help to discriminate between Balmer breaks and line emission, though can sometimes be inconclusive (see text).}
\label{fig:flares_mediumband}
\end{figure}

To further highlight the potential difficulty in distinguishing between line emitters and Balmer-break galaxies (given the available data), we show the SEDs and photometry for two galaxies from the FLARES simulations in Fig.~\ref{fig:flares_mediumband}. The FLARES simulations \citep{Lovell2021, Vijayan2021, Wilkins2023} are a suite of hydrodynamic simulations of galaxy formation and evolution, which use the physics of EAGLE \citep{Crain2015, Schaye2015} and self-consistently models nebular line and continuum emission, as well as the effect of the 3D star--dust geometry on the overall attenuation. These two galaxies were selected to have very similar NIRCam wide-band photometry (dark blue and red circles below 5~\textmu m), yet starkly different origins for the F444W excess, namely $z=8$ \OIII + \Hb\ line emission vs.\@ a $z=10$ Balmer break, respectively. Note that the magnitude difference $\Delta m$ on the vertical axis refers to the magnitude offset relative to F356W, i.e.\@ $\Delta m = m_{F356W} - m_{\lambda}$. Thus it would in principle be difficult to distinguish between these two scenarios, given only the NIRCam wide-band data. The only possible disciminating factor would be the flux in the F115W filter, which is non-zero for the $z=8$ line emitter, but is consistent with zero for the $z=10$ galaxy. However, if the observed data has limited depth, or the rest-frame UV dust attenuation is particularly strong in this band, then the SEDs in practice appear to be identical.

The key differentiating factor between these two SEDs is revealed with the addition of medium-band NIRCam photometry (shown in light blue and orange squares): F410M, F430M, F460M and F480M. Here the signature of the $z=10$ Balmer-break galaxy is revealed through a consistently elevated flux density across the four medium-band filters, while the nature of the $z=8$ line emitter becomes apparent through the high and low flux densities measured by the medium-band filters, which are boosted by line emission and tracing the continuum level, respectively. MIRI F560W imaging (the dark blue and red circles above 5~\textmu m) can in principle also help to discriminate between Balmer breaks and line emitters. However, the main caveat here is that this is not guaranteed. Depending on the ionisation parameter $U$ and metallicity, which control the ratio between the \OIII\ (in F444W) and \Ha\ (in F560W) equivalent width, the \Ha -boosted flux density in the F560W filter can be comparable to that in the \OIII -boosted F444W, thus mimicking the signature of a Balmer break.

\begin{figure*}
\centering
\includegraphics[width=\linewidth]{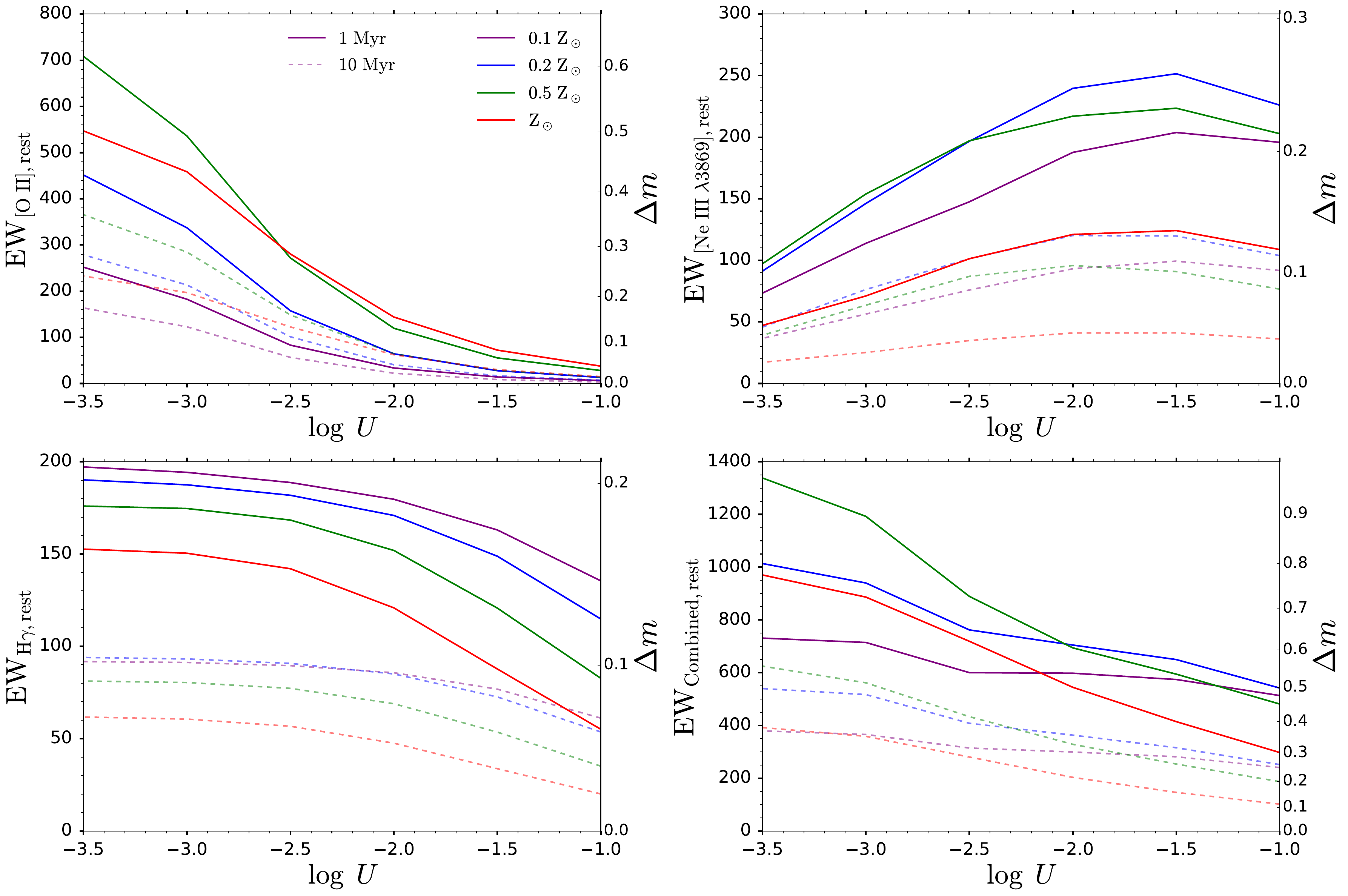}
\caption{The rest-frame equivalent widths for various weak rest-frame optical emission lines, as a function of ionisation parameter $U$, shown both 1~Myr (solid) and 10~Myr (dashed) after an instantaneous starburst, for a range of metallicities (various colours). We use the \citet{Nakajima2022} spectral models for galaxies. Top-left panel: \OII\ doublet. Top-right panel: \NeIII\ $\lambda$3869. Bottom-left panel: \Hg. Bottom-right panel: The combined equivalent width of all of the weak emission lines (\OII, \NeIII, \Hg, \Hd\ and the rest of the fainter Balmer series). Although these lines may individually be weak, when combined together, these lines can have a non-negligible effect on broadband photometry, producing a photometric excess $\Delta m$ = $2.5\log _{10}(1 + \mathrm{EW_{rest}}(1+z)/\Delta \lambda)$ (right axes) that is comparable to our colour criterion for selecting Balmer-break candidates (0.2--0.4~mag). Additional medium-band imaging will be essential to disentangle the contributions of these weak rest-frame optical emssion lines from that of the Balmer break.}
\label{fig:weak_emission_lines}
\end{figure*}

Thus medium-band imaging with all four of the F410M, F430M, F460M and F480M filters, which collectively span the entirety of the F444W filter range, is essential to confidently distinguish between Balmer breaks and strong line emission. Moreover, the inclusion of these medium-bands enables one to disentangle the contributions from line emission and continuum breaks, thus alleviating the need to apply redshift cuts (such as those in this work) to isolate the contributions from Balmer breaks to broadband photometry \citep[see also][]{Laporte2023}. Indeed F410M imaging has already been taken for many of the fields studied in this work. Additional imaging with all three of F430M, F460M and F480M \citep[such as for the JEMS survey,][]{Williams2023} would take roughly half the time per pointing than MIRI F560W. This shorter exposure time, combined with the fact that the NIRCam imaging footprint ($\sim$9.7~arcmin$^2$) is roughly 4$\times$ the MIRI footprint ($\sim$2.35~arcmin$^{2}$) solidifies follow-up medium-band imaging as the ideal next step for identifying Balmer-break galaxies with \emph{JWST}.

Indeed, we note that follow-up spectroscopy with the NIRSpec G395M grating could very rapidly (i.e.\@ with an exposure time comparable to a single medium-band NIRCam filter) identify strong \OIII $\lambda$5007 emission (if there), thus ruling out the Balmer break scenario. However, a lack of strong line emission (due to \OIII\ and \Hb\ being redshifted out) need not necessarily imply the F444W excess (or F356W excess) seen is attributable to a Balmer break, as we will now discuss in the following two sections. 

\subsection{Cumulative weak line emission}

Having discussed the strong emission lines (\OIII\ and \Hb), we now shift our attention to the weaker (i.e.\@ lower equivalent width) emission lines in the rest-frame optical (e.g.\@ \OII, \NeIII, \Hg\ etc.\@). The main point we wish to make in this section is that, although these lines may (perhaps) be individually weak, when combined together, these lines can have a non-negligible contribution to the photometric excess seen in broadband photometry, thus mimicking the Balmer break signature and hence reducing the accuracy with which the Balmer break strength can be determined. 

We show this more clearly in Fig.~\ref{fig:weak_emission_lines}, where we show the rest-frame equivalent width for various weak rest-frame optical emission lines, against ionisation parameter $U$, both 1~Myr (solid) and 10~Myr (dashed) after an instantaneous starburst, for a range of metallicities (various colours). We use the \citet{Nakajima2022} spectral models for galaxies, which incorporate both stellar emission as well as the contributions from nebular line and continuum emission \citep[incorporated using CLOUDY,][]{Ferland1998, Ferland2013}. We note that we obtain similar qualitative results when using the \citet{Zackrisson2011} models and default stellar+nebular Bagpipes templates.

We show the \OII\ doublet equivalent widths in the top-left panel of Fig.~\ref{fig:weak_emission_lines}. The \OII\ equivalent widths can in principle be very high ($\sim$500~\AA\ rest-frame), provided that the galaxy is observed shortly after a starburst, the metallicity is high ($\geq 0.2~\mathrm{Z}_\odot$) and the ionisation parameter is low ($\log U \leq -3.0$). However, while such starburst activity may be likely in the EoR \citep[see e.g.][]{Endsley2023, Looser2023, Looser2023b}, with the young stellar populations potentially outshining the older underlying starlight \citep[see e.g.\@][]{Narayanan2023}, galaxies at these high-redshifts are expected (and indeed have been found) to be metal-poor \citep{Maiolino2008, Maiolino2019, Curti2023, Nakajima2023} and to have relatively high ionisation parameters \citep[see e.g.\@][]{Cameron2023, Curti2023, Matthee2023}. Hence the \OII\ equivalent widths may, in practice, still be relatively low. However, unless there is photometry directly probing the \OIII\ strength (such as F410M and F444W for our $z \sim 8$ sample), this cannot be definitively ruled out from the photometric data alone. At the very least, this \OII\ emission can have a non-negligible contribution to the photometric excess $\Delta m = 2.5\log _{10}(1 + \mathrm{EW_{rest}}(1+z)/\Delta \lambda$) observed in either F444W or F356W, as shown on the right axis of the figure. Here we assume a source at $z=10$, with $\Delta \lambda$ representing the width of the F444W filter $\approx$ 10000~\AA.

Furthermore, as shown in the top-right panel, the \NeIII\ doublet can also contribute to the photometric excess seen in broadband photometry, with the equivalent width increasing with increasing ionisation parameter. Shortly after a starburst, these equivalent widths are $\sim$150~\AA\ rest-frame, producing a photometric excess $\Delta m \sim 0.15$~mag. The \Hg\ equivalent widths and photometric excesses (shown in the bottom-left panel) are comparable to this. Indeed, when combined with the emission from the rest of the Balmer series (i.e.\@ \Hd, \He, H8, etc.\@), the combined Balmer rest-frame equivalent width can be 300--400~\AA, i.e.\@ $\Delta m \sim 0.3$. 

Thus, when combining the contributions from all of these emission lines together (i.e.\@ \OII, \NeIII, \Hg\ and the rest of the Balmer series, see bottom-right panel), the net equivalent width and thus effect on broadband photometry can be considerable, generating a photometric excess $\Delta m$ that is at the very least comparable to our colour criterion for selecting strict/relaxed Balmer-break candidates (0.2--0.4~mag). See also the elevated flux density in F356W (compared to F277W) for the $z=8$ line emitter in Fig.~\ref{fig:flares_mediumband} for a visual SED example of this. Thus it can be difficult to distinguish between a Balmer break and the cumulative effect of these weak emission lines, using wide-band NIRCam photometry alone.

Hence, additional medium-band imaging (F410M, F430M, F460M and F480M) will be essential to definitively disentangle the contributions of these weak rest-frame optical emission lines from that of the Balmer break (at $z \sim 10.5)$. Indeed, these four medium-band filters collectively span the width of the F444W filter, thus enabling one to isolate the (slight) emission from individual lines from the underlying continuum level. At $z \sim 8$, the (F300M, F335M, F360M, F410M) filter set achieve the same result for the F356W filter. 

\subsection{Dusty continuum or a (dusty) AGN}

In addition to line emission, a steeply rising continuum level in the rest-frame optical can mimic the photometric excess signature of a Balmer break. As has been discussed in the literature, this steeply rising continuum level can be driven by a dusty continuum \citep[see e.g.\@][]{Laporte2021} or a (dusty) AGN \citep[see][]{Furtak2022, Barro2023, Kocevski2023, Labbe2023b, Matthee2023b}. We briefly explore these two scenarios in this section.

Firstly, we note that our Balmer-break candidates tend to have relatively flat (or at most slightly red) SEDs in the rest-frame UV (see Fig.~\ref{fig:f444w_candidates} and Fig.~\ref{fig:f356w_candidates}), with the SED suddenly jumping in flux density in the rest-frame optical, due to the (supposed) Balmer break. Hence this SED profile is inconsistent with a single component galaxy that is affected by dust attenuation, as we would expect (given the wavelength dependence of  dust attenuation laws) the attenuation to be strongest in the rest-frame UV and weaker in the optical (i.e.\@ the opposite of what is seen in the data). Hence we assume that our galaxy consists of two components: a relatively flat component (in terms of $f_\nu$) that dominates in the rest-frame UV, and a second, dust-reddened component that begins to dominate in the rest-frame optical. If the photometric excess seen at $z \sim 10.5$ in the F444W band is driven by a steeply rising dusty continuum level, and we assume that all of the measured F356W and F444W flux originates from this dusty emission, then our $> 0.4$~mag colour requirement for Balmer-break candidates translates into a minimum E(B$-$V) of 0.37, i.e.\@ $A_\mathrm{V}$ $\approx$ 1.5 (assuming a \citealt{Calzetti2000} dust attenuation law). Although galaxies at these high-redshifts are not expected to be particularly dust-rich \citep[with $A_\mathrm{1600} \approx 2.5 A_\mathrm{V} \approx 1$ for bright $z\sim 7 $ Lyman break galaxies, see][]{Bowler2018, Bowler2022}, at the very least a dusty continuum can partially contribute to the photometric excess seen in the F444W (or F356W) bands. 

\begin{figure}
\centering
\includegraphics[width=\linewidth]{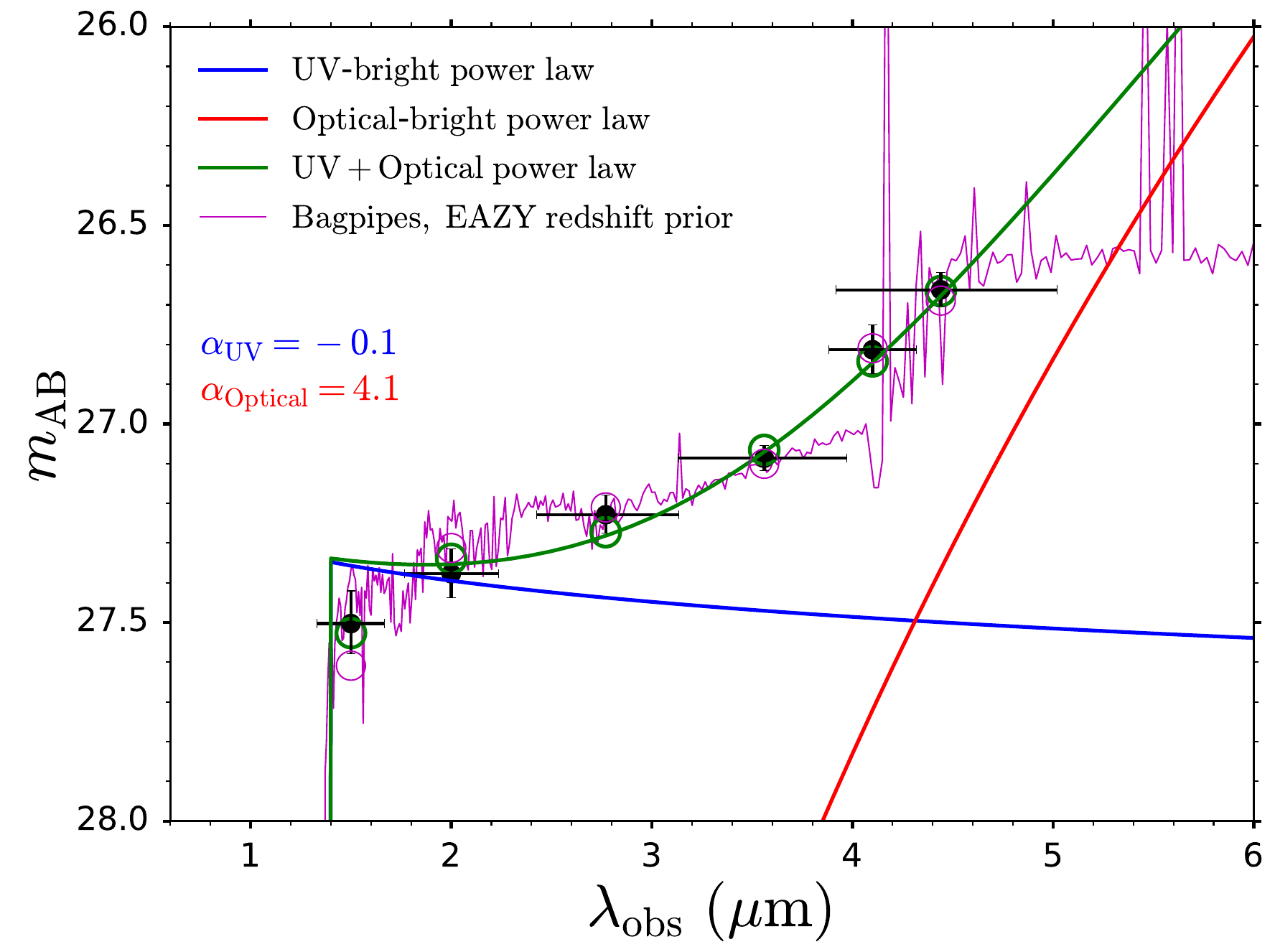}
\caption{Double power law fit (green) to the NIRCam photometry (black) of a $z \sim 10.5$ strict Balmer-break candidate. The relatively flat/blue power law (shown in blue, with $\alpha = -0.1$) that dominates in the rest-frame UV represents a young/unobscured component, while the (very) red power law (shown in red, with $\alpha = 4.1$) that dominates in the rest-frame optical represents a dusty galaxy or (dusty) AGN component. The double power law fit provides a reasonable match to the photometry for this particular Balmer-break candidate (though this is not the case for all of our candidates), yielding a comparable quality of fit to the stellar+nebular Bagpipes result (purple). Thus a steeply rising dusty/AGN continuum level in the rest-frame optical can in principle mimic (or at least contribute to) the photometric excess signature of a Balmer break, making it difficult to properly distinguish between these scenarios with NIRCam wide-band imaging alone. As before, additional medium-band imaging (as well as MIRI imaging) enables one to break this degeneracy, as these filters would trace this rapidly rising continuum (if present).}
\label{fig:power_law_fit}
\end{figure}

Alternatively, the photometric excess seen may be driven by the steeply rising continuum level of a (dusty) AGN component, which begins to dominate the SED in the rest-frame optical. Assuming that this AGN emission follows a power law, with $f_\nu \propto 
\nu^{-\alpha}$, and is contributing all of the measured F356W and F444W flux, then our $> 0.4$~mag colour requirement for Balmer-break candidates translates into a rest-frame optical power law index requirement of $\alpha_\mathrm{optical} > 1.67$. In practice, the minimum power law slope will likely have to be even steeper than this, due to the UV-bright component still contributing a non-negligible (or even substantial) amount of flux to the F356W and F444W bands (see Fig.~\ref{fig:power_law_fit}). We note that in the \citet{Nakajima2022} models, AGN spectra with UV power law indices of [$1.2$, $1.6$ and $2.0$] are considered. Although the power law slope tends to be less steep (i.e.\@ a smaller $\alpha$) in the rest-frame optical (which is what our photometric excess is probing), with the inclusion of additional dust attenuation the $\alpha_\mathrm{optical} > 1.67$ requirement seems within reach of the expectations for AGN models.

To highlight this further, we show an example of a double component galaxy fit in Fig.\@~\ref{fig:power_law_fit}, with a blue/flat power law component that dominates in the rest-frame UV (shown in blue) and a red power law component that begins to dominate in the rest-frame optical (red). Combined (shown in green), this double power law fit provides a reasonable match to the data for this particular $z{\sim}10.5$ strict Balmer-break candidate (though this is not the case for all of our candidates), yielding a comparable quality of fit to the stellar+nebular Bagpipes result. We note that the inferred power law slope for the optically-bright component is very red, with $\alpha = 4.1$, which is comparable to the typical optical slopes inferred for extremely red dusty galaxy/AGN candidates at $z\sim7.5$ \citep{Barro2023}.

\begin{figure*}
\centering
\includegraphics[width=\linewidth]{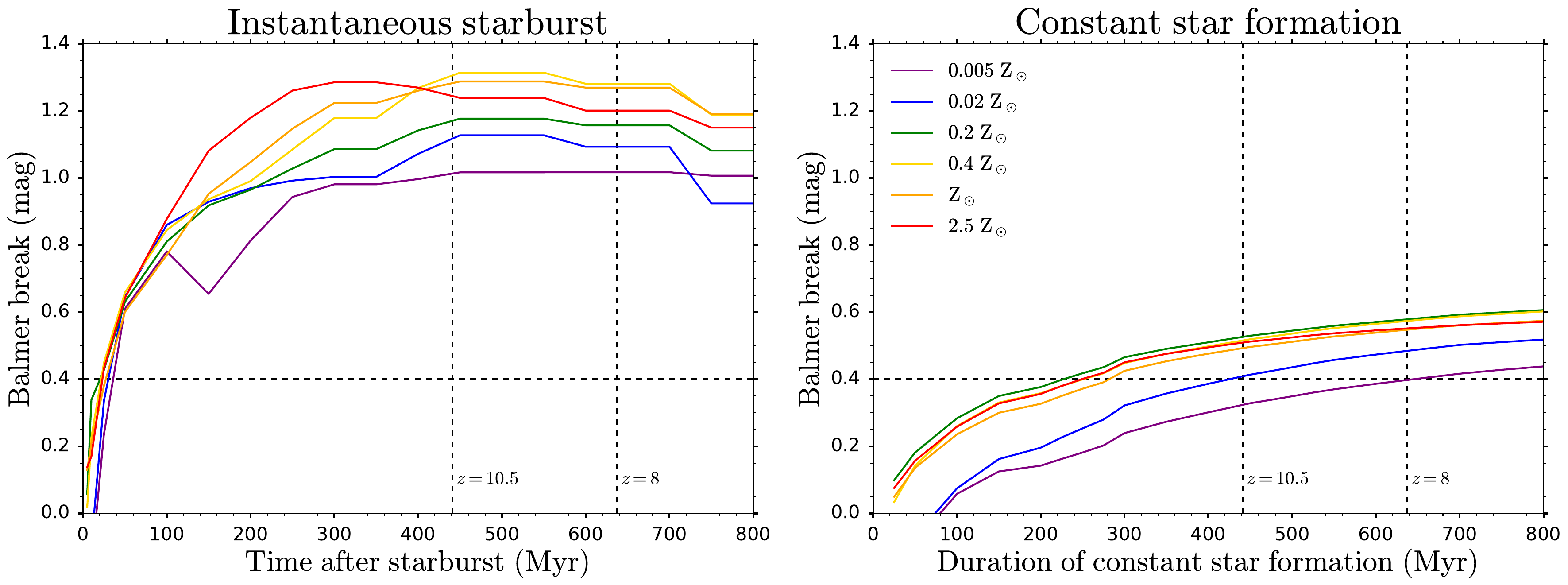}
\caption{Balmer break strength (in magnitudes, see text for definition) against time, for a range of metallicities (various colours). We use the default stellar and nebular templates from the Bagpipes SED-fitting code and assume zero dust attenuation. Left panel: The evolution of the Balmer break strength following an instantaneous starburst. Right panel: The time-evolution of the Balmer break strength during a constant star-formation history. Depending on the star-formation history, the Balmer break strength either rapidly ($<50$~Myr, left) or very slowly ($>250$~Myr, right) passes our 0.4~mag colour selection threshold (horizontal dashed lines) . For reference, the vertical dashed lines refer to the age of the Universe at $z=8$ and $z=10.5$. Thus the Balmer break strength is very dependent on the star-formation history. A 0.4~mag Balmer break therefore does not necessarily imply a particularly old stellar population. Likewise, an old stellar population (at these redshifts) may not necessarily exhibit a large Balmer break. Hence the Balmer break strength alone cannot provide definitive constraints on the stellar age of the galaxy. Deep NIRSpec continuum spectroscopy, together with MIRI imaging, will ultimately provide the best possible SFH contraints, thus also providing the best possible indirect constraints on the onset of star formation in the Universe.}
\label{fig:balmer_break_sb_cont}
\end{figure*}

Thus it is at the very least plausible that a dusty continuum or a (dusty) AGN can contribute (somewhat) to the photometric excess seen in the F444W or F356W bands. Indeed, with just the NIRCam wide-band imaging alone, there is likely insufficient information to properly distinguish between these two scenarios and a Balmer break \citep[see also][]{Barro2023}. As before, additional medium-band imaging again enables one to break this degeneracy and place tighter constraints on the Balmer break strength. In contrast to a Balmer break, where the continuum level redward of the break is relatively flat, the continuum level in the two aforementioned scenarios rises much more steeply, which can readily be identified using the (F410M, F430M, F460M, F480M) filter set, which would trace this rapidly rising continuum level (if present). In a similar vein, longer wavelength imaging with e.g.\@ MIRI F560W would also aid in the identification of dust/AGN, again due to the steeper SEDs expected in these scenarios. However, as outlined before, the medium-band imaging is both more time- and area-efficient, and thus remains the ideal follow-up strategy also in this scenario. 

\subsection{The Balmer break strength as a proxy for stellar age}\label{subsec:break_caveats}

Even if both strong and weak line emission can be ruled out (or constrained), as well as the contributions from a dusty continuum and/or (dusty) AGN, the Balmer break strength alone likely cannot provide definitive constraints on the stellar ages of galaxies \citep[see also][]{Wilkins2023}. This is because the Balmer break strength depends on the full star-formation history of the galaxy, which can be complex, given the likely bursty SFHs of galaxies in the EoR \citep[see e.g.\@][]{Endsley2023, Looser2023, Looser2023b}.

We show this more clearly in Fig.~\ref{fig:balmer_break_sb_cont}, where we show how the Balmer break strength (in magnitudes, following our definition in Section~\ref{sec:results}) evolves with time, for a range of different metallicities (various colours). For this analysis we use the default stellar and nebular templates from the SED-fitting code Bagpipes, and assume zero dust attenuation.

In the left panel of Fig.~\ref{fig:balmer_break_sb_cont}, we see that the Balmer break strength rapidly rises following an instantaneous starburst, passing our 0.4~mag colour criterion in less than 50~Myr. Hence a moderately-sized Balmer break in this scenario does not imply particularly old stars, nor does it therefore place particularly strong constraints on the onset of star formation in the Universe in this case. On the other hand, we see that the Balmer break strength rises much more slowly in the case of a constant star-formation history (right panel). In the more metal-enriched scenarios ($\geq 0.2~\mathrm{Z}_\odot$), it takes $\sim$250~Myr of constant star formation to pass our 0.4~mag threshold, and even takes up to the age of the Universe at $z=10.5$ and $z=8$ (vertical dashed lines) at even lower metallicity. Hence a moderately-sized Balmer break implies very old stellar ages in this scenario, and would therefore place very strong indirect constraints on the onset of star formation in the Universe in this case. 

Furthermore, we note that, from our investigations, a 1~Myr starburst is able to mask out the Balmer break signature of e.g.\@ a 200~Myr old starburst, even if the young population only has 3\% the mass of the older stars. Hence the Balmer break strength is very dependent on the star-formation history of the galaxy.

Thus we wish to make three key summarising points regarding the use of the Balmer break strength as a proxy for the presence of old stars in high-redshift galaxies. Firstly, that a moderately-sized Balmer break (such as the 0.4~mag for our colour criterion) need not imply a particularly old stellar population. Secondly, that a galaxy with old stellar populations does not necessarily have a particularly strong Balmer break. Thirdly, that it is only the strongest Balmer breaks (perhaps 0.8~mag or so) that likely imply the presence of relatively old stars (taking e.g.\@ $\sim$100~Myr after an instantaneous starburst and unachievable within the age of the Universe at these redshifts for a constant SFH). However, such large Balmer breaks in the observed data should be interpreted with caution, as they may be attributable to strong line emission (\OIII\ and \Hb) that is misidentified as a Balmer break \citep[see also][]{Adams2023, Trussler2023}.

Thus the Balmer break strength alone cannot provide definitive constraints on the age of the stellar population. Indeed, \citet{Wilkins2023}, who investigate the Balmer break strength in the FLARES simulations in greater detail, arrive to a similar conclusion. However, as we have seen in Section~\ref{sec:results} and will discuss in more detail in the next two subsections, a galaxy with a larger Balmer break does, on average, have an older stellar age. As outlined earlier, additional medium-band imaging will be essential to robustly identify Balmer-break galaxies. These Balmer-break galaxies can then be followed-up with deep continuum spectroscopy with the NIRSpec PRISM, together with MIRI imaging, to place the strongest possible constraints on their star formation histories, thus providing the best possible indirect constraints on the onset of star formation in the Universe. Owing to the fact that the NIRSpec PRISM starts to quickly drop off in sensitivity above 3.56 \textmu m, these integrations will take an order of magnitude longer to achieve 5$\sigma$ detections compared to the triad of F430M, F460M and F480M imaging that preceded it.

\subsection{Applying our colour selection to the FLARES simulations}

Having discussed the caveats for identifying Balmer breaks from photometric data, together with the constraints these breaks can place on the stellar ages of these galaxies, we now apply our Balmer break colour selection to the FLARES simulations to investigate the properties of the galaxies selected in this manner.

\begin{figure}
\centering
\includegraphics[width=.875\linewidth]{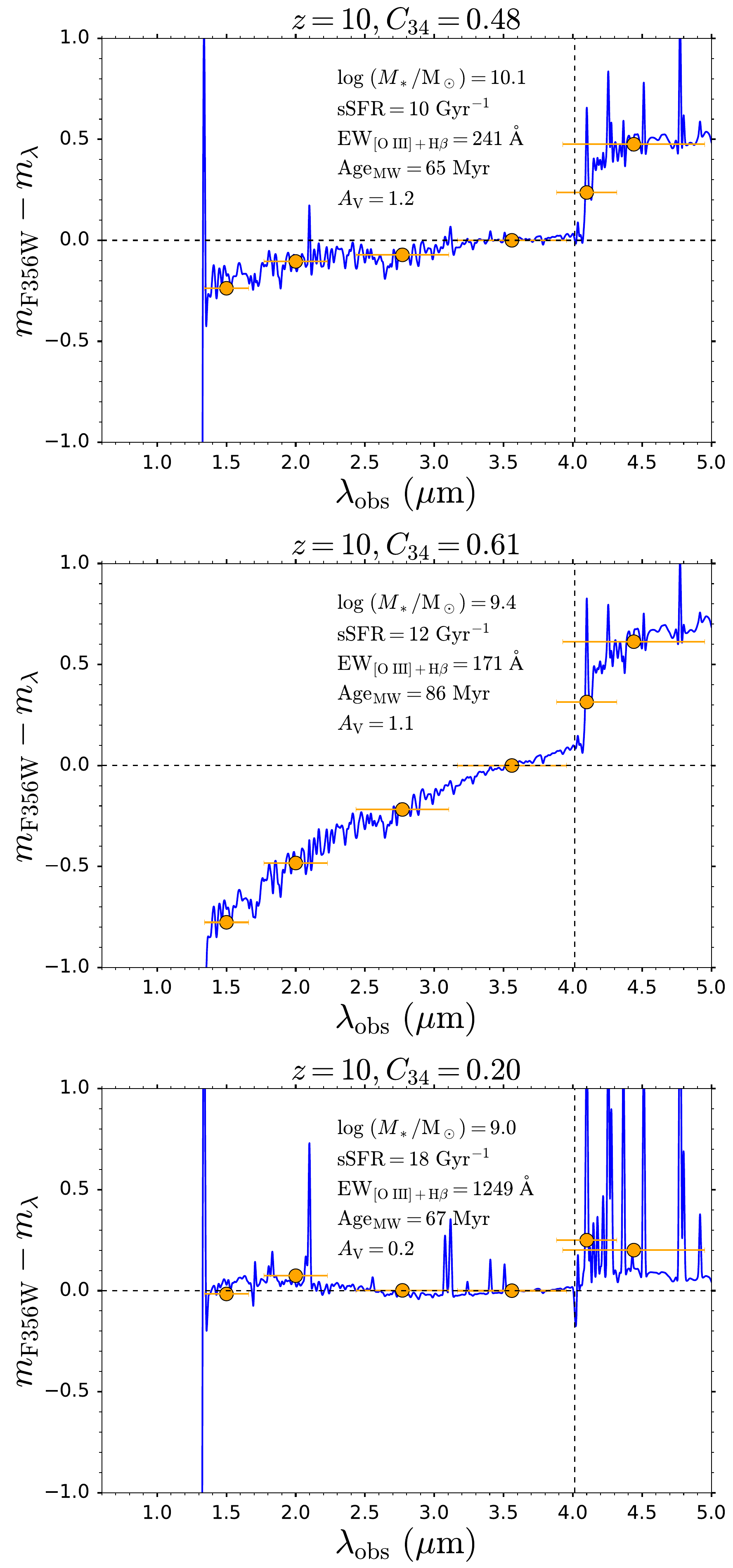}
\caption{SEDs (blue) and NIRCam photometry (orange) for three $z=10$ galaxies from the FLARES simulations. Here we show $m_\mathrm{F356W} - m_\lambda$, i.e.\@ the magnitude difference with F356W, to highlight the strength of the Balmer break (position indicated by the vertical dashed line). The F356W$-$F444W colour $C_{34}$ is denoted at the top of each panel, and the stellar mass, specific star formation rate, \OIII\ + \Hb\ combined rest-frame equivalent width, mass-weighted stellar age and V-band attenuation $A_{V}$ are denoted within the panel. Top panel: A galaxy with a Balmer break and relatively flat rest-frame UV SED that is selected by our strict colour criterion. Middle panel: A galaxy with a Balmer break and red rest-frame UV SED that is selected by our strict colour criterion. Bottom panel: A galaxy without a prominent Balmer break, but with weak line emission, that would be selected if the colour criterion is reduced.}
\label{fig:flares_examples}
\end{figure}

The FLARES simulations outputs are provided at redshifts of $z= 5, 6, 7, 8, 9, 10$. Here we focus on our selection of $z=10$ galaxies through the F444W excess colour criterion (i.e.\@ F356W$-$F444W $> 0.4$). We show the example SEDs and photometry for three FLARES galaxies (and also list some of their properties, i.e.\@ stellar mass, specific star formation rate, \OIII\ + \Hb\ rest-frame equivalent width, mass-weighted stellar age and V-band attenuation) in Fig.~\ref{fig:flares_examples}. Here we show the magnitude difference relative to F356W (i.e.\@ $m_\mathrm{F356W} - m_\lambda$), to highlight the strength of the Balmer break. We find that with our F356W$-$F444W $> 0.4$~mag criterion we exclusively select galaxies with Balmer breaks at $z = 10$ (though we note that AGN emission is not included in the FLARES SEDs). We select galaxies both with a relatively flat UV SED (top panel), as well as galaxies with red UV SEDs (middle). Furthermore, if we relax the F444W excess criterion, such as for our redshift-dependent colour selection, we begin to select galaxies without prominent Balmer breaks, where the F444W excess is driven by the cumulative effect of the weak rest-frame optical emission lines (bottom panel).  

We note that the fraction of galaxies with F356W$-$F444W $>$ 0.4 increases with increasing stellar mass at $z=10$. While only 15\% of galaxies with $8.0 < \log (M_*/\mathrm{M}_\odot) < 9.0$ satisfy this criterion, this increases to 34\% at $9.0 < \log (M_*/\mathrm{M}_\odot) < 9.5$ and 56\% at $9.5 < \log (M_*/\mathrm{M}_\odot) < 10.0$. 

Additionally, the $\log (M_*/\mathrm{M}_\odot) > 8$, $z=10$ galaxies that satisfy our Balmer break colour criterion tend to have weaker emission lines, with median \OIII\ $\lambda$5007,  \Hb, \NeIII\ $\lambda$3869 and \OII\ doublet rest-frame equivalent widths of 77~\AA, 25~\AA, 3~\AA\ and 5~\AA\ respectively, compared to the non-selected galaxies, which have 463~\AA, 118~\AA, 17~\AA\ and 18~\AA. These Balmer break colour-selected galaxies also tend to have stronger Balmer breaks (0.58~mag vs.\@ 0.19~mag) and tend to be older, with median mass-weighted stellar ages of 94~Myr vs.\@ the 58~Myr for the non-selected galaxies.

\subsection{Additional colour selections}

\begin{figure}
\centering
\includegraphics[width=.875\linewidth]{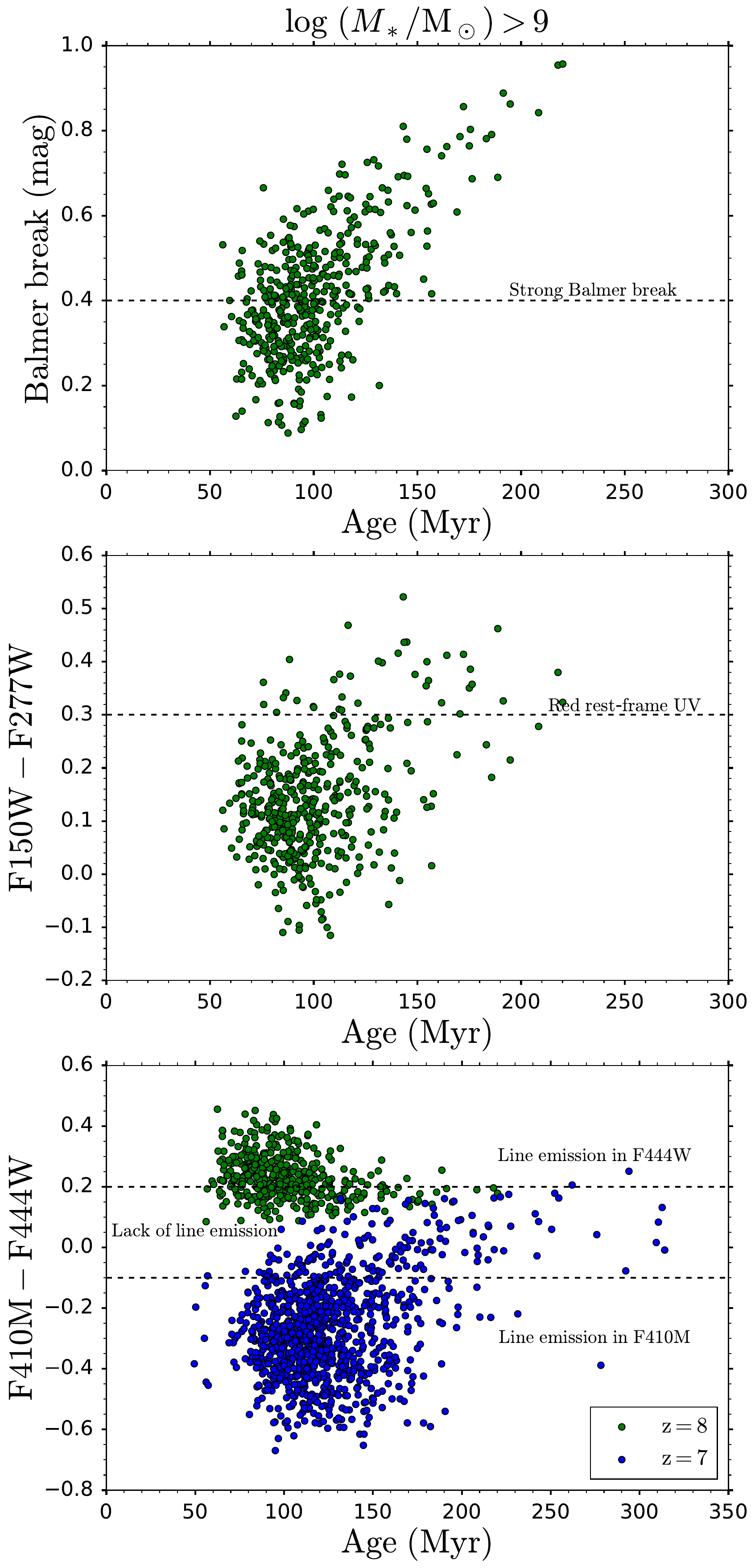}
\caption{Various colour selections (horizontal dashed lines) applied to $z=8$ (green) and $z=7$ (blue) galaxies from the FLARES simulations. The mass-weighted stellar age is displayed on the horizontal axis. Only galaxies with stellar masses $\log (M_*/\mathrm{M}_\odot) > 9$ are shown. Top panel: Balmer break selection. Middle panel: Rest-frame UV colour selection, as traced by F150W$-$F277W. Bottom panel: F410M$-$F444W colour selection, a proxy for the current star formation activity for galaxies at $z=$ 7--9. Flat (i.e.\@ close to zero) F410M$-$F444W colours likely indicate a lack of line emission (from \OIII\ and/or \Hb) and thus low sSFR at these redshifts, which is what drives the preferential selection of older galaxies. In general the galaxy stellar age does not correlate well with the Balmer break strength or rest-frame UV colour, except for the largest breaks (which motivates the primary colour selection adopted in this paper) and reddest UV colours, for which the ages tend to be older than from a random selection.}
\label{fig:flares_selection}
\end{figure}

We have seen that galaxies with moderately-sized Balmer breaks do not necessarily have particularly old stellar ages. However, what is true is that, on average, galaxies with moderately-sized Balmer breaks will tend to be older than galaxies without such breaks. Here we use the FLARES simulations to explore additional colour cuts (beyond the Balmer break selection) that can be applied to increase the likelihood of selecting a galaxy with old stars (though at the cost of completeness). 

Here we consider two additional colour cuts. The first involves the rest-frame UV colour F150W$-$F277W. The second involves the F410M$-$F444W colour, a proxy for current star formation activity. As such we restrict our FLARES analysis to galaxies at $z=8$ (and $z=7$), where the F410M and F444W filters probe the strong rest-frame optical emission lines. 

The basis behind the rest-frame UV colour cut is that, on average, galaxies with redder F150W$-$F277W colours will tend to be older, as old galaxies have red rest-frame UV slopes. Thus we require a red rest-frame UV colour, which we deem to be (from a visual inspection of Fig.~\ref{fig:flares_selection}) F150W$-$F277W $>$ 0.3. Of course, dusty star-forming galaxies, which have young stellar ages, can also have red slopes and would be selected in this way. On the other hand, old galaxies with a young blue starbursting component that dominates their rest-frame UV would not be selected. 

The principle behind the F410M$-$F444W colour cut is that galaxies that are currently not actively star-forming (i.e.\@ with a low sSFR) will tend to be older. Thus the F410M and F444W filters, which are redward of the Balmer break at $z=8$ (and $z=7,9$) should have comparable flux densities, as they both probe the continuum level. In other words, the F410M$-$F444W colour should be close to zero. Crucially, these two filters also cover the strong \OIII\ and \Hb\ lines at these redshifts, and owing to the fact that the F410M filter is narrower than the F444W filter, it is more sensitive to and thus more strongly boosted by line emission. Hence at $z=7$, where the \OIII\ line is both in the F410M and F444W filters, any ongoing (strong) star formation would manifest itself as a blue F410M$-$F444W colour. Similarly, at $z=8$, where \OIII\ is still in the F444W filter (but not in F410M), any ongoing star formation would result in a red F410M$-$F444W colour. Thus our F410M$-$F444W colour cut requires this colour to be close to zero, to rule out strong line emission, i.e.\@ $-0.1 < $ F410M$-$F444W $<$ 0.2. However, similar to the caveats with the previous colour selection, galaxies with no ongoing star formation need not necessarily be old. Furthermore, galaxies with old stars can still currently be actively star-forming. 

We show these two colour selections, together with our Balmer break colour selection, against the mass-weighted stellar ages of FLARES galaxies in Fig.~\ref{fig:flares_selection}. We note that only galaxies with stellar masses $\log (M_*/\mathrm{M}_\odot) > 9$ are shown in this diagram. 

In the top panel of Fig.~\ref{fig:flares_selection}, we find that, as has been discussed before, that the Balmer break strength does not correlate very well with the stellar age, provided that the Balmer break strength is small. It is only for the largest Balmer break values, that a larger break strength generally is associated with an older stellar population. 

Similarly, in the middle panel of Fig.~\ref{fig:flares_selection}, the rest-frame UV colour does not generally correlate very well with the galaxy age. It is only for the reddest, most extreme colours, that the age tends to be higher than for a random selection, thus motivating this additional colour selection. 

Finally, in the bottom panel of Fig.~\ref{fig:flares_selection}, we show our F410M$-$F444W colour selection. As can be seen, the $z=7$ galaxies with flat (i.e.\@ close to zero) colours and presumably a lack of line emission tend to be older than the rest of the population. Likewise, the older $z=8$ galaxies tend to have more flat F410M$-$F444W colours, though the trends are clearly not as strong as for $z=7$.

We remark that these additional colour selections appear to become less effective at lower stellar masses ($8 < \log (M_*/\mathrm{M}_\odot) < 9$). These lower mass galaxies tend to have bluer UV slopes, with now only a single galaxy satisfying our UV colour requirement. 

Thus ideally (and ultimately) the full photometry should be used to infer the presence of evolved stars in high-redshift galaxies. However, medium-band imaging is essential for the reliable identification of Balmer breaks and thus for placing definitive constraints on the stellar populations in these galaxies. In the absence of such data, applying single (or multiple) colour cuts increases the likelihood of selecting galaxies with evolved stellar populations. 

\section{Conclusions} \label{sec:conclusions}

In this work we utilised a combination of PEARLS GTO and public NIRCam photometric data to search for Balmer-break candidate galaxies at $7 < z < 12$. In principle, the detection of a Balmer break implies the presence of evolved stars in these high-redshift galaxies, thus placing valuable indirect constraints on the onset of star formation in the Universe. 

As the signature of a Balmer break in photometry is a photometric excess seen in the rest-frame optical, i.e.\@ the longer wavelength NIRCam bands, we use the measured F444W excess or F356W excess as the basis for our identification of Balmer-break candidates at $z\sim 10.5$ and $z \sim 8$, respectively. We investigate and compare the physical properties and star formation histories of our Balmer-break candidates against a control sample of galaxies (i.e.\@ galaxies at the same redshifts which do not exhibit a F444W or F356W excess).

Regarding the magnitudes (i.e.\@ brightnesses) and stellar masses of our Balmer-break candidates, we find the following: 
\begin{itemize}
\item Our Balmer-break candidates are similarly distributed in brightness compared to the non-Balmer break control sample.
\item However, we do find that our Balmer-break candidates tend to be more massive for a given brightness than the control sample. 
\item Therefore, our results suggest that the (very) large stellar masses inferred for some red rest-frame optical galaxies in the EoR \citep[$\log\, (M_*/\mathrm{M}_\odot) > 10$,][]{Labbe2023}, may be driven by their higher inferred mass-to-light ratios (a result of the SED-fitting process and assumptions therein), rather than these galaxies being inherently ultra-bright.
\item We find that strong line emission, the cumulative effect of weak emission lines, a dusty continuum and a (dusty) AGN can all contribute to the photometric excess seen in broadband photometry. This somewhat alleviates the need for evolved stellar populations, bringing the inferred stellar mass-to-light ratios down, thus likely lowering the (extremely) large stellar masses inferred for some high-redshift galaxies to be more in line with the predictions from theoretical simulations of galaxy formation.
\end{itemize}

Furthermore, utilising the Bagpipes SED-fitting code to derive constraints on the stellar ages, star formation histories, emission line equivalent widths, specific star formation rates and UV slopes $\beta$, we find the following: 
\begin{itemize}
\item Our $z \sim 10.5$ sample of Balmer-break candidates (i.e.\@ galaxies with a F444W excess) tend to be older (with a median age of 115~Myr), have more extended star formation histories, have lower inferred \OIII\ + \Hb\ equivalent widths (120~\AA\ rest-frame), have lower specific star formation rates (6~Gyr$^{-1}$) and redder UV slopes ($\beta = -1.8$) than the control sample of galaxies. This suggests that the observed strength of (what we believe to be) a Balmer break places tight constraints on the age and star-formation history of a galaxy.
\item However, these trends all become less strong for our $z \sim 8$ sample of Balmer-break candidates. Now it is only true that, on average, the Balmer-break candidates display older ages (100~Myr) than the control sample (30~Myr), i.e.\@ a Balmer break alone does not guarantee an old stellar population.
\item Indeed, the F444W filter now probes the strong rest-frame optical \OIII\ and \Hb\ emission lines in these $z~\sim 8$ galaxies, thus providing valuable additional constraints on their current star formation activity.
\item Thus longer wavelength photometry which probes beyond the Balmer break, such as MIRI F560W, will likely be essential to place stronger constraints on the star formation histories on the highest redshift galaxies, such as our $z \sim 10.5$ sample.
\item The weakened connection between the (supposed) observed Balmer break strength (i.e.\@ F356W excess) and the inferred galaxy stellar age may be attributable to the (likely) bursty nature of these EoR galaxies, causing more of a disconnect between their current star formation activity and SED profiles, and their more extended star-formation history.
\end{itemize}

Finally, we discuss caveats regarding the identification of Balmer breaks from photometric data, as well as the constraints these Balmer breaks can place on the presence of evolved stellar populations in high-redshift galaxies. We find the following:
\begin{itemize}
\item Despite applying redshift cuts to remove strong \OIII\ + \Hb\ line emitters from our Balmer-break candidate sample, such line emitters can still be confused for Balmer-break galaxies due to the limited constraining power of wide-band photometric data, thus resulting in inaccurate photometric redshifts.
\item The cumulative emission from the weak rest-frame optical emission lines (\OII, \NeIII, \Hg, \Hd\ and the remainder of the fainter Balmer series) can contribute a non-negligible amount of flux density to broadband filters (with a photometric excess $\Delta m$ that can be a few tenths of a mag), thus mimicking the photometric excess signature of a Balmer break. 
\item Both a dusty continuum, as well as a (dusty) AGN, which begin to dominate in the rest-frame optical, can contribute to the photometric excess seen in broadband filters. 
\item All of the above processes can therefore (somewhat) masquerade as a Balmer break in the wide-band photometric data, thus reducing the accuracy with which the Balmer break strength can be inferred. 
\item Additional medium-band imaging, which can clearly discriminate between these scenarios, will therefore be essential for the reliable identification of Balmer breaks. At $z \sim 10.5$ the (F410M, F430M, F460M, F480M) filter set collectively span the full width of the F444W filter, thus enabling one to disentangle the contributions from line emission, a sharply rising dust/AGN continuum and a Balmer break. The (F300M, F335M, F360M, F410M) filter set achieves the same purpose at $z \sim 8$.
\item Even if these other processes can be ruled out or accounted for, the Balmer break strength alone cannot serve as the definitive indicator of the stellar age of the galaxy. This is because it exhibits a complex dependence on the star-formation history of the galaxy.
\end{itemize}
Thus, ultimately, deep NIRSpec continuum spectroscopy, together with MIRI imaging, will be needed to provide the strongest possible constraints on the star formation histories of these high-redshift galaxies. In turn, these observations will place the strongest indirect constraints on the onset of star formation in the Universe, thereby revealing when cosmic dawn breaks.

\section*{Acknowledgements}

We dedicate this paper to the memory of our dear PEARLS colleague Mario Nonino, who was a gifted and dedicated scientist, and a generous person.
We thank the referee for their useful comments which helped to improve this article. JT, CC, NA and QL acknowledge support from the ERC Advanced Investigator Grant EPOCHS (788113). DA and TH acknowledge support from STFC in the form of PhD studentships. LF acknowledges financial support from Coordenação de Aperfeiçoamento de Pessoal de Nível Superior - Brazil (CAPES) in the form of a PhD studentship. R.W., S.C., and R.J.\@
acknowledge support from NASA JWST Interdisciplinary Scientist grants NAG5 12460, NNX14AN10G and 80NSSC18K0200 from GSFC. MAM acknowledges the support of a National Research Council of Canada Plaskett Fellowship, and the Australian Research Council Centre of Excellence for All Sky Astrophysics in 3 Dimensions (ASTRO 3D), through
project number CE17010001. CNAW acknowledges funding from the JWST/NIRCam contract NASS-0215 to the University of Arizona, as well as support from the NIRCam Development Contract NAS5-02105 from NASA Goddard Space Flight Center to the University of Arizona. M.N. acknowledges INAF-Mainstreams 1.05.01.86.20. APV acknowledges support from the Carlsberg Foundation (grant no CF20-0534). The Cosmic Dawn Center (DAWN) is funded by the Danish National Research Foundation under grant No.140. 

This work is based on observations made with the NASA/ESA \emph{Hubble Space Telescope} (\emph{HST}) and NASA/ESA/CSA \emph{James Webb Space Telescope} (\emph{JWST}) obtained from the Mikulski Archive for Space Telescopes (MAST) at the Space Telescope Science Institute (STScI), which is operated by the Association of Universities for Research in Astronomy, Inc., under NASA contract NAS 5-03127 for \emph{JWST}, and NAS 5–26555 for \emph{HST}. The observations used in this work are associated with \emph{JWST} programs 1176 and 2738. In addition, public datasets from \emph{JWST} programs 1324 (GLASS), 1345
(CEERS) and 2079 (NGDEEP) are also used within the work presented.

This research made use of Astropy,\footnote{http://www.astropy.org} a community-developed core Python package for Astronomy \citep{astropy2013, astropy2018}.

%%%%%%%%%%%%%%%%%%%%%%%%%%%%%%%%%%%%%%%%%%%%%%%%%%
\section*{Data Availability}

The \citet{Nakajima2022} models used in this article will be shared on reasonable request to Kimihiko Nakajima. The FLARES simulation results are publicly available at: https://flaresimulations.github.io/. The catalogues and imaging used in this publication will be made publicly available once initial works are completed with students involved in the data reduction and analysis. Products using GTO data will be made available as and when exclusive access periods lapse. Any remaining data underlying the analysis in this article will be shared on reasonable request to the first author.



\bibliographystyle{mnras}
\bibliography{main.bib} % if your bibtex file is called example.bib

%\begin{thebibliography}{99}
%\end{thebibliography}

%%%%%%%%%%%%%%%%%%%%%%%%%%%%%%%%%%%%%%%%%%%%%%%%%%

%%%%%%%%%%%%%%%%% APPENDICES %%%%%%%%%%%%%%%%%%%%%

\appendix

%%%%%%%%%%%%%%%%%%%%%%%%%%%%%%%%%%%%%%%%%%%%%%%%%%

% Don't change these lines
\bsp	% typesetting comment
\label{lastpage}
\end{document}